\documentclass[10pt,a4paper]{article} 
\usepackage[utf8]{inputenc}
\usepackage[english]{babel}
\usepackage{amsmath}
\usepackage{amsfonts}
\usepackage{amssymb}

\usepackage[backend=bibtex, style=numeric-comp]{biblatex}
\bibliography{quellen.bib}

\usepackage{mathtools} 
\usepackage{graphicx}
\usepackage{subfigure}
\usepackage{epstopdf}
\usepackage{amsthm}        
\usepackage{hyperref} 
\usepackage{multicol} 
\usepackage{microtype} 
\usepackage{booktabs} 
\usepackage{float} 
\usepackage{tikz} 
\usetikzlibrary{matrix, arrows}
\usepackage{tabularx}
\usepackage{longtable} 
\usepackage{array}
\usepackage[hmarginratio=1:1,top=20mm,left=25mm, right=25mm, columnsep=20pt]{geometry} 
\usepackage[hang, small,labelfont=bf,up,textfont=it,up]{caption} 

\usepackage[hang]{footmisc} 
\usepackage{lettrine} 
\usepackage{paralist} 

\usepackage{abstract} 

\usepackage{titlesec} 
\renewcommand\thesection{\Roman{section}} 
\renewcommand\thesubsection{\Roman{section}.\arabic{subsection}} 
\titleformat{\section}[block]{\bf\large\scshape\centering}{\thesection.}{1em}{} 
\titleformat{\subsection}[block]{\bf\large}{\thesubsection.}{1em}{} 

\newtheoremstyle{defi+ex} 
                        {0.5cm}    
                        {0.5cm}    
                        {\it}         
                        {}         
                        {\bfseries}
                        {}        
                        {\newline} 
                        {\normalfont \bfseries{\thmname{#1}\thmnumber{ #2}.\hspace{2mm}\thmnote{(#3)}}}         

\newtheoremstyle{rem} 
                        {0.5cm}    
                        {0.5cm}    
                        {\it}         
                        {}         
                        {\bfseries}
                        {}        
                        {0mm} 
                        {\normalfont \bfseries{\textit{\thmname{#1}\thmnumber{ #2}.\hspace{2mm}\thmnote{(#3)}}}}         

\theoremstyle{defi+ex}
\newtheorem{defi}{Definition}
\newtheorem{ex}{Example}
\newtheorem{prop}{Proposition}

\theoremstyle{rem}
\newtheorem{rem}{Remark}[section]

\title{\vspace{-15mm}\fontsize{24pt}{10pt}\selectfont\textbf{The signed permutation group on Feynman graphs}} 

\author{
\large
\textsc{Julian Purkart}
\\[2mm] 
\normalsize Institute of Physics, Humboldt University  \\ 
\normalsize Newtonstr. 15, D-12489 Berlin, Germany   \\ 
\normalsize \href{mailto:purkart@physik.hu-berlin.de}{purkart@physik.hu-berlin.de} 
}
\date{\today}

\begin{document}

\maketitle

\begin{abstract}
\noindent 
The Feynman rules assign to every graph an integral which can be written as a function of a scaling parameter $L$. 
Assuming $L$ for the process under consideration is very small, so that contributions to the renormalizaton group are small,  we can expand the integral and only consider the lowest orders in the scaling. The aim of this article is to determine specific combinations of graphs in a scalar quantum field theory that lead to a remarkable simplification of the first non-trivial term in the perturbation series.  
It will be seen that the result is independent of the renormalization scheme and the scattering angles.
To achieve that goal we will utilize the parametric representation of scalar Feynman integrals as well as the Hopf algebraic structure of the Feynman graphs under consideration. Moreover, we will present a formula which reduces the effort of determining the first-order term in the perturbation series for the specific combination of graphs to a minimum.
\end{abstract}

\begin{multicols*}{2} 

\section{Introduction}

In physics, the probability amplitude of an interaction process between elementary particles can be calculated as a perturbation series in the scaling. The coefficients in the perturbative expansion of the correlation function (or Green's function) are integrals which can be interpreted as physical processes. Graphically, these processes can be represented via Feynman diagrams, which are the central objects of perturbative quantum field theory. To treat them in an adequate manner, it will be necessary to get familiar with some fundamental aspects and definitions of graph theory which will be covered in section \ref{subsec:graph theoretic foundations}.  Afterwards, in section \ref{subsec:graph polynomials},  we introduce polynomials associated with the respective graphs, the first and second Symanzik polynomial. In section \ref{sec:parametric renorm}, these polynomials will also show up in the integrand 
of the parametric Feynman integral which can be obtained by going from momentum to parametric space, using the so-called Schwinger trick. At the end of section \ref{sec:preliminaries} we establish an algebraic structure on the set of Feynman graphs and thereby give a brief insight in the Hopf algebra of rooted trees. \\
As already mentioned, Feynman diagrams are connected with the integrals in the perturbation series. 
This connection is given by the Feynman rules under which each graph is mapped to an integral. 
The problem arising from these rules is that the resulting integrals are not ensured to be convergent and well-defined. Indeed, plenty of them are divergent and therefore we have to renormalize the integrals.
%
In section \ref{sec:parametric renorm}, this will be done for a scalar quantum field theory. 
Therefore, we rescale the integral whereby it can be written as a function of a scale $S$ and dimensionless scattering angles $\Theta$. 
Applying kinetic renormalization conditions to the integral and using the forest-formula indicates that the renormalized Feynman-rules can be written as a polynomial in the scaling parameter $L$ (cf. \cite{BroKr:AnglesScales}).\\
The starting point of section \ref{sec:L-linear term} is to give the notion of flags. Afterwards, the $L$-linear term of the renormalized Feynman rules is considered for antisymmetric flags, which are sums over permutations $\sigma \in S^{\mbox{\tiny signed}}_{n-1} \times S^{\mbox{\tiny cyclic}}_{n}$ of nested graph insertions: with the result that it is independent of the scattering angles.\\
The results of the preceding section are used in section \ref{sec:general formula} to provide a general formula that allows to compute the $L$-linear term of the renormalized Feynman rules for antisymmetric flags regardless of the number of graphs inserted into each other.
This formula is based on the idea of finding all partitions of a graph's rooted tree instead of its forest-set. A pictorial approach to manage this task is given in section \ref{sec:pictorial approach}, inspired by the notion of Ferrers diagrams (see appendix \ref{app:ferrers diagram}).\\
The article is concluded by an example in which the $L$-linear term of the renormalized Feynman rules is computed for an antisymmetric flag of co-radical degree six.

\section{Preliminaries} \label{sec:preliminaries}
The definitions in subsection \ref{subsec:graph theoretic foundations} basically follow \cite{BroKr:AnglesScales}, \cite{KrSarsSuij:Gauge} \cite{Suij:RenormGaugeField}, and \cite{West:GraphTheory}. 
The principal sources of subsection \ref{subsec:graph polynomials} are \cite{BogWein:GraphPoly} and \cite{BroKr:AnglesScales}. 
The subsection \ref{subsec:hopf algebra} as well as the appendix \ref{app:algebras} is based on  \cite{BergKr:HopfAlgebra}, \cite{EbraKr:HopfAlg}, \cite{Kass:QuantumGroups}, and \cite{Manchon:HopfAlgebra}.\\

\subsection{Graph theoretical foundations}\label{subsec:graph theoretic foundations}

\begin{figure}[H]
	\begin{center}
$\mathcal R_{QED} = \left\{
\vcenter{\hbox{\includegraphics[scale=0.15]{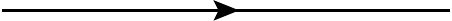}}}, 
\vcenter{\hbox{\includegraphics[scale=0.15]{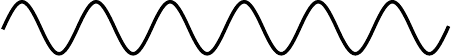}}}, 
\vcenter{\hbox{\includegraphics[scale=0.15]{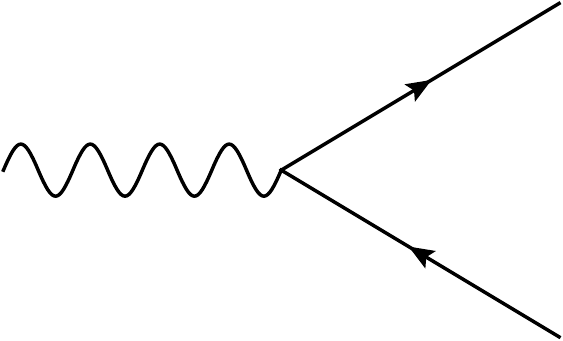}}}
\right\}$ \\
$\mathcal R_{QCD} = \left\{
\vcenter{\hbox{\includegraphics[scale=0.13]{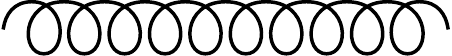}}}, 
\vcenter{\hbox{\includegraphics[scale=0.13]{graphs/fermion}}}, 
\vcenter{\hbox{\includegraphics[scale=0.13]{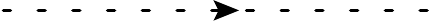}}}, 
\vcenter{\hbox{\includegraphics[scale=0.13]{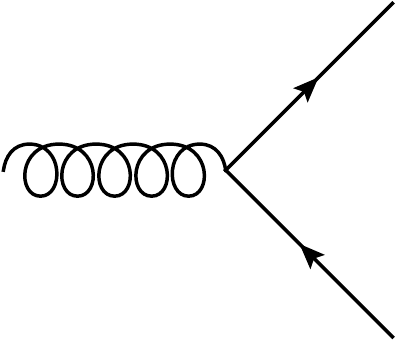}}}, 
\vcenter{\hbox{\includegraphics[scale=0.13]{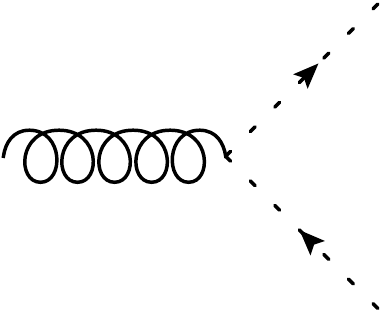}}}, 
\vcenter{\hbox{\includegraphics[scale=0.13]{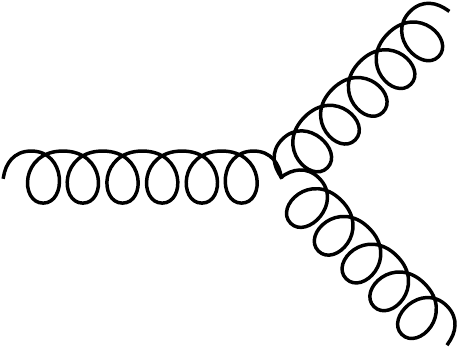}}}, 
\vcenter{\hbox{\includegraphics[scale=0.13]{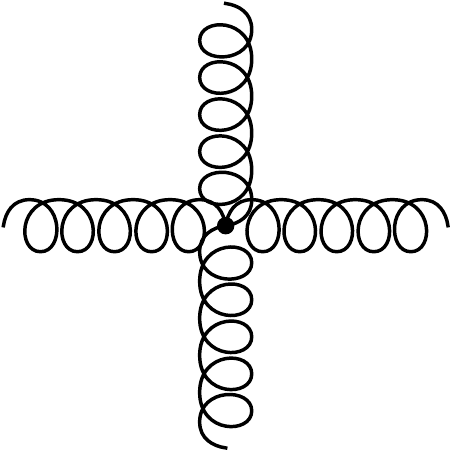}}}
\right\}$ \\ 
$\mathcal R_{\phi^4} = \left\{
\vcenter{\hbox{\includegraphics[scale=0.175]{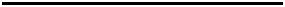}}},
\vcenter{\hbox{\includegraphics[scale=0.175]{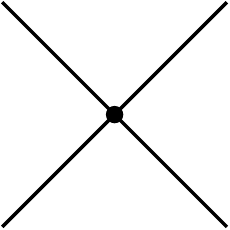}}}
\right\}$
	\end{center}
\caption{Sets of allowed vertices and edges for QED, QCD, and $\phi^4$.}
\label{fig:set vertices and edges}
\end{figure}

In this section, we want to acquaint ourselves with some basic definitions of graph theory and a bit of vocabulary needed when talking about Feynman graphs and graphs in general. Therefore, we first have to define:
\begin{defi}[Graphs] \label{def:graph}
A graph $G = \left(E, V, \Phi\right)$ consists of a set of edges $E$, a set of vertices $V$, and a map $\Phi$ (incidence relation) from edges to pairs of vertices.
	\begin{itemize}
	\item An edge $e\in E$ is said to be incident to $v,w\in V$ if $ \Phi(e)=\{v,w\}$. The vertices $v,w$ are called the endpoints of $e$. 
	\item Two vertices $v,w\in V$ are adjacent if $\exists e\in E$ such that $\Phi(e) = \{v,w\}$.
	 \item The valence of a vertex is given by the number of edges incident to it.
	 \item All edges are oriented meaning that each edge $e \in E$ directs from a source vertex $s(e) \in V$ to a target vertex $t(e) \in V$.
	 \item An edge with equal endpoints, i.e. $s(e) = t(e)$ is called a loop.
	 \item A path $(v,w)$ of length $k$ from $v$ to $w$ is given by a subset $E_P = \left\{e_1,\dots,e_k\right\}\subseteq E$ such that any two edges $e_i, e_{i+1}$ have at least one endpoint in common. If $v = w$, the path is called a cycle.
	 \item If there exists a path $(v,w)$ for any pair $v,w\in V$, the graph is called connected. A connected graph without loops is said to be simply connected.
	
	\end{itemize}
\end{defi}
In addition to the graph theoretical definition given above, there are some features that come up when treating Feynman graphs instead of standard graphs. In general, the edges and vertices of a Feynman graph are labeled, that is assigning information of physical interest to them like the momentum and mass of the particles. Furthermore, Feynman graphs are constructed from a particular set of edges and vertices which we will denote by $\mathcal R = \mathcal R_E \cup \mathcal R_V$, following \cite{Kr:AnatomyGauge} and \cite{Suij:RenormGaugeField}. While $\mathcal R_E$ corresponds to the type of quantum particles, $\mathcal R_V$ determines the type of interaction between those particles, respectively. Generally, the sets $\mathcal R_E$ and $\mathcal R_V$ are dictated and restricted by the quantum field theory we are looking at. In some theories the edges also get an orientation, corresponding to the charge flow of the particles. 
 In figure \ref{fig:set vertices and edges} the sets of vertices and edges are given for quantum electrodynamics (QED), quantum chromodynamics (QCD), and $\phi^4$-theory
\footnote{$\phi^k$-theories are scalar field theories treating only one kind of particles with spin zero	 represented by the one-component scalar field $\phi$. Those particles self-interact in groups of $k$ which means that all vertices are $k$-valent.}
in $D=4$ dimensions of space-time. \\
In the following we will denote such Feynman graphs by $\Gamma$ with vertex set $\Gamma^{[0]}$ and edge set $\Gamma^{[1]}$. In contrast to standard graph theory, we have to distinguish between internal and external edges. An edge is called internal if it connects two vertices whereas an external edge connects only to one vertex, that is to say it has only one endpoint. The set of edges then is given by the union $\Gamma^{[1]} = \Gamma_{\mbox{\tiny int}}^{[1]} \cup \Gamma_{\mbox{\tiny ext}}^{[1]}$.
\begin{defi}[Feynman graphs] \label{def:feynman graph}
A Feynman graph $\Gamma = \left(G, \operatorname{res}\right)$ is given by a graph $G$ and a map $\operatorname{res}$
\begin{align}
\operatorname{res} : \Gamma^{[0]} \cup \Gamma^{[1]} \rightarrow \mathcal R_V \cup \mathcal R_E
\end{align}
which assigns to each vertex and edge in $\Gamma$ an element from a set of allowed types of edges and vertices. The elements $r\in \mathcal R$ are called the allowed residues of the theory.\\
For any connected Feynman graph $\Gamma$ we let $\operatorname{res}\left(\Gamma\right)$ be the graph $\Gamma$ when all its internal edges shrink to one point. Then, $\operatorname{res}\left(\Gamma\right)$ is just the residue of the graph, defining its external structure.
\end{defi}
The allowed residues of a theory form the set of building blocks such that each Feynman graph of the theory can be built up out of it.
In the literature, the terms Feynman graph/diagram and graph/diagram are often used interchangeably and so will we do in the following. Moreover, we will only consider a special kind of graphs, called one-particle irreducible graphs, throughout this article.
\begin{defi}[One-particle irreducible graphs]\label{def:1PI graph}
A connected Feynman graph $\Gamma$ is said to be one-particle irreducible (1PI) if it is still connected after removing one of its internal edges. Depending on the number of external edges, there are several kinds of 1PI graphs:
	\begin{itemize}
	\item If $\Gamma$ has no external edges, it is called a vacuum graph or vacuum bubble.
	\item For $|\Gamma_{\mbox{\tiny ext}}^{[1]}| = 1$ the graph is called tadpole.
	\item If $|\Gamma_{\mbox{\tiny ext}}^{[1]}| = 2$, we call $\Gamma$ a propagator or self-energy graph.
	\item All other graphs with $|\Gamma_{\mbox{\tiny ext}}^{[1]}| \geq 3$ are said to be vertex graphs.
	\end{itemize}
\end{defi}

\begin{figure}[H]
	\begin{center}
$\Gamma = \vcenter{\hbox{\includegraphics[scale=0.2]{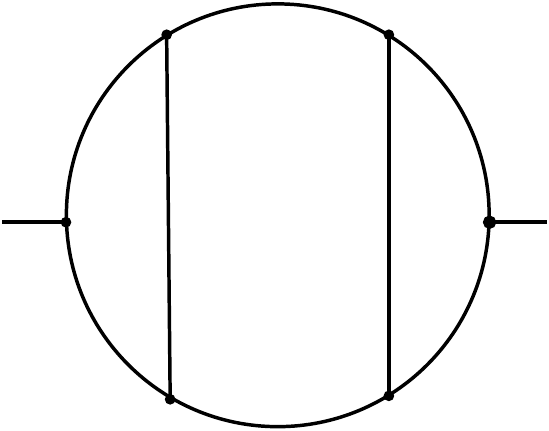}}}  \hspace{5mm}
\gamma = \vcenter{\hbox{\includegraphics[scale=0.2]{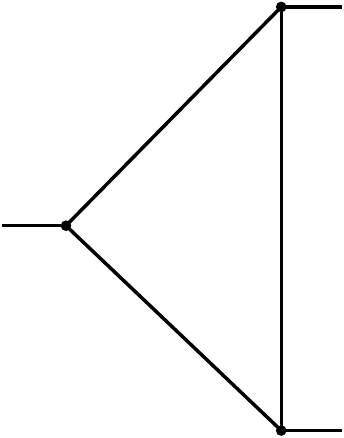}}} \hspace{5mm}
\Gamma / \gamma = \vcenter{\hbox{\includegraphics[scale=0.2]{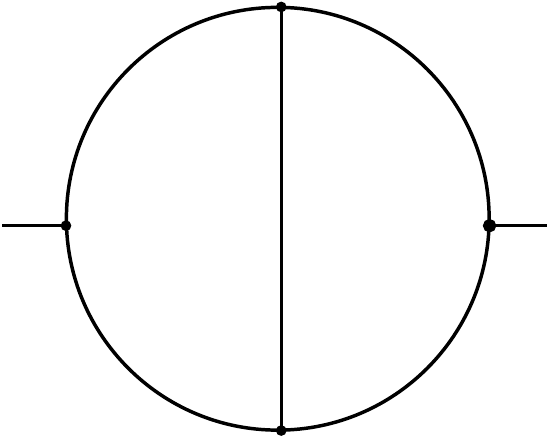}}}
$
	\end{center}
\caption{Example for a subgraph $\gamma$ and the corresponding cograph $\Gamma / \gamma$ of the three-loop graph $\Gamma$.}
\label{fig:bsp sub/cograph}
\end{figure}

\begin{defi}[Sub- and cographs] \label{def:sub-cograph}
A graph $\gamma \subseteq \Gamma$ is called subgraph of $\Gamma$ if $\gamma^{[0]} \subseteq \Gamma^{[0]}$, $\gamma^{[1]} \subseteq \Gamma^{[1]}$, and the assignment of endpoints to edges in $\gamma$ and $\Gamma$ is the same. In the case that $\gamma$ contains all vertices of $\Gamma$, i.e. $\gamma^{[0]} = \Gamma^{[0]}$, $\gamma$ is said to be a spanning subgraph of $\Gamma$. \\
The cograph $\Gamma / \gamma$ is obtained from $\Gamma$ by shrinking all internal edges of $\gamma$ in $\Gamma$ to length zero, i.e. to a single point, such that the external leg structure is not affected, $\operatorname{res}\left(\Gamma/\gamma\right) = \operatorname{res}\left(\Gamma\right)$. The operation "$/$", which can be used to reverse graph insertions, is called contraction. Using this notion, the map $\operatorname{res}$ acting on a connected graph $\Gamma$, can be seen as the maximal contraction $\Gamma/\Gamma$.
\end{defi}
An example for a sub- and cograph is given in figure \ref{fig:bsp sub/cograph}.
Note that contracting a $2$-point (also propagator or self-energy) graph leads to two different kinds of $2$-point vertices related to the mass and the momentum of the propagating particle. A detailed discussion on this topic is given in \cite{CoKr:RHProb1}.\\
Before the Feynman graph polynomials are invented in the next subsection, we will need two special types of graphs.

\begin{defi}[Tree] \label{def:tree} 
A connected and simply connected (no cycles) graph is called a tree $T$ with vertex set $T^{[0]}$ and edge set $T^{[1]}$.
	\begin{itemize}
	\item A rooted tree is a tree $T$ with a distinguished vertex $r\in T^{[0]}$, which is called the root, such that all edges are oriented away from it.
	\item The weight $|T|$ of a tree is given by its number of vertices.
	\item Let $\mathcal{T}_r$ be the set of all rooted trees and $\mathcal{T}_r^{(i)}$ the subset of all rooted trees with weight $|T| = i$, $\forall$ $ T \in \mathcal{T}_r^{(i)}$, then we can write $\mathcal{T}_r= \bigcup_i \mathcal{T}_r^{(i)}$.
	\item A rooted tree is said to be decorated if there exists a finite set $D$ of decorations and a surjective map $c: D \rightarrow T^{[0]}$, which assigns to each vertex $v\in T^{[0]}$ an element $d\in D$.
	\end{itemize}
\end{defi}

\begin{defi}[Forest
\footnote{It should be pointed out that there are two different definitions of the notion of a forest. In the present case we define the forest (of subdivergences) in the context of renormalization and Hopf algebra. This definition is also in accordance with the forest formula introduced in section \ref{sec:parametric renorm}. Within the framework of graph polynomials (cf. subsection \ref{subsec:graph polynomials}) the forest (or $k$-forest) is defined as a graph without cycles/loops consisting of $k$ connected components. That is, a $k$-forest is given by the disjoint union of $k$ trees. For example, the forest set in figure \ref{fig:bsp spanning 2 forest} follows this definition.
}
] \label{def:forest}
Let $\Gamma$ be a Feynman graph and $f \coloneqq \left\{\gamma_i\right\}$ a subset of divergent 1PI proper subgraphs $\gamma_i \subsetneq \Gamma$ such that for any $\gamma, \gamma' \in f$ one of the following conditions is fulfilled:
\begin{align}
\gamma \subset \gamma' , \hspace{2mm} \gamma' \subset \gamma ,  \mbox{\hspace{2mm}or\hspace{2mm}} \gamma \cap \gamma' = \emptyset.
\end{align}
That is, the elements of $f$ are either disjoint or contained in each other. Then, $f$ is called a forest and $\mathcal{F}\left(\Gamma\right)$ denotes the set of all forests of the graph.
	\begin{itemize}
	\item A forest $f$ of a Feynman graph $\Gamma$ is said to be maximal if and only if the cograph $\Gamma/f = \Gamma/\cup_{\gamma\in f} \gamma$ is a 1PI graph, not containing any divergent proper 1PI subgraphs. Such graphs are called primitive.
	\item A maximal forest $f$ of $\Gamma$ is complete if any $\gamma\in f$ is either primitive or there exists a proper subgraph $\gamma'\in f$ of $\gamma$ such that the cograph $\gamma/\gamma'$ is primitive.
	\item If $f$ consists of $k$ connected components, it is called a $k$-forest. A $1$-forest is a tree. 
	\item The union of rooted trees gives a rooted forest and its set is denoted by $\mathcal{F}_r$.
	\end{itemize}
\end{defi}

\begin{figure}[H]
	\begin{center}
$
\vcenter{\hbox{\includegraphics[scale=0.4]{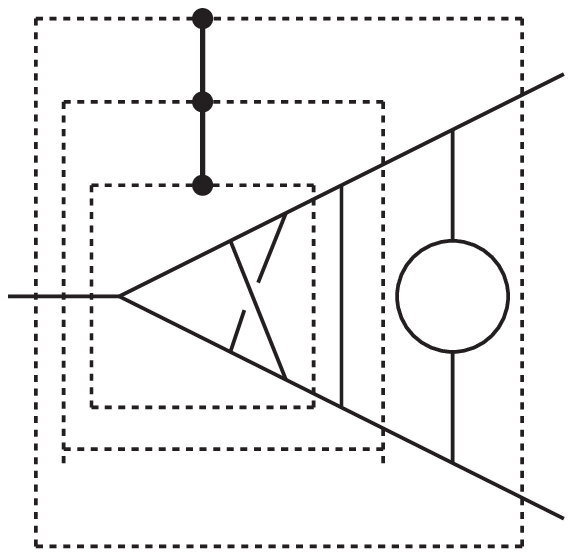}}} \Longleftrightarrow
\vcenter{\hbox{\includegraphics[scale=0.235]{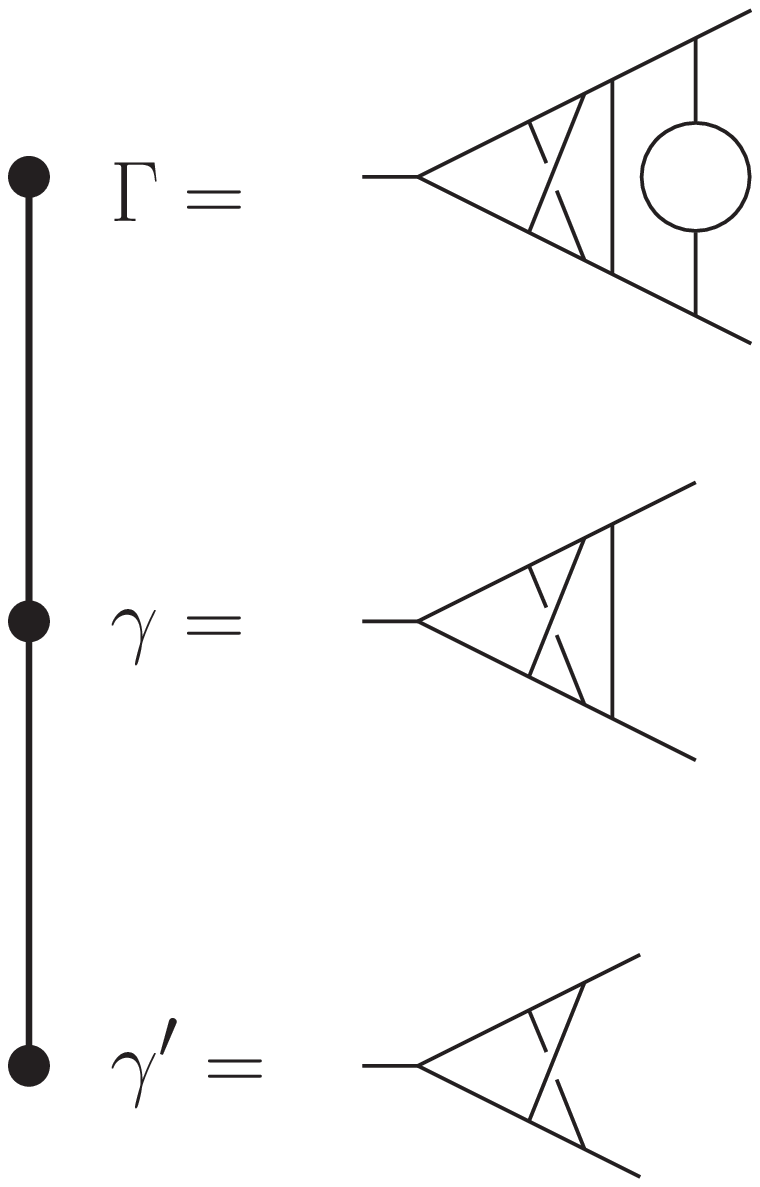}}}
$
	\end{center}
\caption{Example from $\phi^3_6$-theory for a rooted tree of a graph associated with its subgraph structure}
\label{fig:rooted tree box}
\end{figure}

Hereafter, we will often restrict ourselves to trees and forests which are spanning subgraphs of the considered graph. In this case, we call them spanning trees and spanning forests respectively. It is also important not to confuse spanning and rooted trees and forests. While the sets of spanning trees and forests of a graph will be used to define the graph polynomials in section \ref{subsec:graph polynomials}, the sets of rooted trees and forests do not correspond to a specific graph even though one or more elements of $\mathcal{T}_r$ can be associated to a graph, representing its subgraph structure, as we will see. Moreover, we will set up a Hopf algebra structure on the set of rooted trees in section \ref{subsec:hopf algebra}.\\
Assume that $f = \left\{\gamma, \gamma'\right\}$ is a complete forest of $\Gamma$ with primitive elements $\gamma/\gamma'$ and $\gamma'$. Then, we can write $f$ as a sequence of subsets
\begin{align}\label{eq:subset complete forest}
\gamma'\subsetneq \gamma\subsetneq \Gamma
\end{align}
to show how the graph and the subgraphs are nested. Using the notion of forests and trees, we can associate a decorated rooted tree to each complete forest of a graph $\Gamma$. Taking the complete forest in equation (\ref{eq:subset complete forest}), the corresponding decorated rooted tree is given by 
\begin{align}
\vcenter{\hbox{\includegraphics[scale=0.5]{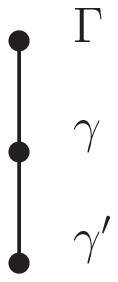}}} 
\hspace{8mm} \mbox{ or } \hspace{15mm} 
\vcenter{\hbox{\includegraphics[scale=0.5]{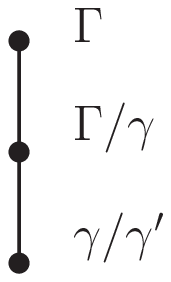}}} .
\end{align}
It becomes apparent that each Feynman diagram $\Gamma$ furnishes a tree whose decorations are the elements of the complete forest. The rooted tree of a graph can also be read off from the box system as one can see in figure \ref{fig:rooted tree box}, in which each box contains a divergent subgraph of the graph and corresponds to a leaf of the tree. The root is given by the whole graph (the outermost box). Like the elements in the complete forest, the boxes are not allowed to overlap, but rather are nested or disjoint. 

%
The case of overlapping subdivergences is treated in \cite{Bor:AlgLatt} where their structure is analyzed using algebraic lattice theory.

\subsection{Feynman graph polynomials}\label{subsec:graph polynomials}
Using the preceding notion of spanning trees and forests, we want to introduce the Feynman graph polynomials. These polynomials, known as the first and second Symanzik polynomial, have many special properties and can be read off directly from the corresponding graph. Also, they play a crucial role in the computation of Feynman loop integrals since they are directly related to the integrand of such integrals. From the variety of methods to determine the graph polynomials, we will only consider one by interpreting the polynomials in terms of spanning trees and spanning forests. Likewise, it is also possible to compute them with the aid of matrices, associated to the graph. This approach suits well when performing computer algebra since, after the particular matrices are known, the only thing left to do is computing the determinant of a matrix. The basic principle of this approach is the matrix-tree theorem, invented by Gustav Kirchhoff, which exhibits the possibility to compute the number of a graph's spanning trees as the determinant of a matrix derived from the graph (see \cite{BogWein:GraphPoly} for example).
In addition to his contributions to the fundamental understanding of electric circuits and spectroscopy, Kirchhoff was also the one who invented the notion of graph polynomials. \\
Moreover, it is also possible to use graph homology instead of the matrix-tree theorem to derive the graph polynomials.\\
Instead of the matrix-tree theorem one can also use graph homology to derive the polynomials. 
Throughout this article let $\Gamma$ be a connected graph with $E_\Gamma \coloneqq \left|\Gamma^{[1]}_{\mbox{\tiny int}}\right|$ internal edges, $V_\Gamma \coloneqq \left|\Gamma^{[0]}\right|$ vertices, and loop number $L\left(\Gamma\right)$ defined by
\begin{align} \label{eq:loopnumber}
L\left(\Gamma\right) = E_\Gamma - V_\Gamma + 1 .
\end{align}
This number is also called the first Betti number or the cyclomatic number of the graph. For disconnected graphs we have to replace $1$ by $k$, with $k$ the number of connected components of the graph.
Furthermore, let $\mathcal{F}_s^{(k)}$ be the set of all spanning $k$-forests (see definition \ref{def:forest}) and $\mathcal{F}_s$ be the set of all spanning forests of the graph $\Gamma$, given by 
\begin{align}
\mathcal{F}_s = \bigcup_k \mathcal{F}_s^{(k)} .
\end{align}
Then $f\in \mathcal{F}_s^{(k)}$ can be obtained from $\Gamma$ by deleting $L + k - 1$ of its internal edges. The elements of a spanning $k$-forest are composed of the connected components $T_i$ of $\mathcal{F}_s^{(k)}$, which are necessary trees, and will be denoted by
\begin{align}
\bigcup_{i=1}^k T_i = \left(T_1, T_2, \dots, T_k\right) \in \mathcal{F}_s^{(k)} .
\end{align}
From now on we will only consider scalar Feynman graphs in $D$ dimensions of spacetime. The edges of the graph are associated with particles of mass $m_e$. The momenta of the particles will be denoted by $p_e$ for $e\in\Gamma_{\mbox{\tiny ext}}^{[1]}$ and $q_e$ for $e\in\Gamma_{\mbox{\tiny int}}^{[1]}$. We impose momentum conservation at each vertex, i.e. the sum of the momenta flowing into the vertex equals the sum of all outgoing momenta. 
Particularly, with regard to the external momenta it follows that $\sum_{e\in\Gamma_{\mbox{\tiny ext}}^{[1]}} p_e = 0$ if the momenta are taken to flow outwards. Therefore, the internal momenta $q_e$ of a tree graph are completely determined by the external momenta $p_e$. To determine the internal momenta of a loop graph uniquely we have to add $L\left(\Gamma\right)$ (see eq. (\ref{eq:loopnumber})) internal momenta $k_j$, with $j$ labeling the independent loops of the graph $\Gamma$. 
Since these loop-momenta correspond to virtual particles that do not show up in the initial or final state we have to integrate over all possible values, and therefore the amplitude does not depend on the $k_j$ after all. \\
The two Symanzik polynomials of a graph $\Gamma$ are defined as follows.
\begin{defi}[First Symanzik polynomial] \label{def:first symanzik polynomial}
Let $\mathcal{F}_s^{(1)}$ be the set of all spanning trees of $\Gamma$, such that $T\in \mathcal{F}_s^{(1)}$ is obtained from $\Gamma$ by deleting $L$ of its internal edges. We introduce parameters $\alpha_e \in \mathbb{R}_+$ associated to the internal edges $e\in \Gamma_{\mbox{\tiny int}}^{[1]}$ of the graph $\Gamma$.
 Then, the first Symanzik polynomial is defined by
\begin{align} \label{eq:first symanzik polynomial}
\psi_\Gamma = \sum_{T \in \mathcal{F}_s^{(1)}} \prod_{e \notin T^{[1]}} \alpha_e
\end{align}
where the sum is over all spanning trees of $\Gamma$ and $T^{[1]}$ denotes the edge set of $T$.
\end{defi}
The parameters $\alpha_e$ are the so-called Schwinger parameters, which will also show up in the parametric representation of Feynman integrals in section \ref{sec:parametric renorm}.
\begin{defi}[Second Symanzik polynomial] \label{def:second symanzik polynomial}
Let $\mathcal{F}_s^{(2)}$ be the set of all spanning 2-forests of $\Gamma$, such that $\left(T_1, T_2\right) \in \mathcal{F}_s^{(2)}$ is obtained from $\Gamma$ by deleting $L + 1$ of its internal edges. The mass of the particles associated with the edges of the graph $\Gamma$ will be denoted by $m_e$. Then, the second Symanzik polynomial is defined by
\begin{align}\label{eq:second symanzik polynomial}
\phi_\Gamma = \varphi_\Gamma + \psi_\Gamma \sum_{e\in\Gamma_{\mbox{\tiny int}}^{[1]}} \alpha_e m_e^2
\end{align}
with
\begin{align}
\varphi_\Gamma = - \hspace{-4mm} \sum_{\left(T_1,T_2\right)\in\mathcal{F}_s^{(2)}} \hspace{-1mm} Q(T_1) \cdot Q(T_2) \prod_{e\notin T_1^{[1]} \cup T_2^{[1]}} \alpha_e .
\end{align}
The sum is over all spanning $2$-forests of the graph $\Gamma$, and $Q(T_i)$ denotes the sum of all euclidean momenta flowing into the tree $T_i$.
\end{defi}
By momentum conservation it is clear that the sum of all incoming momenta of $T_i$ and the sum of all momenta flowing outwards only differ in the sign. Therefore, the product $Q(T_1) \cdot Q(T_2)$ is equal to minus the square of the sum of the momenta flowing through the cut lines from one tree to the other. Obviously, it is generally valid that $Q(T_1) = - Q(T_2)$ and hence $Q(T_1) \cdot Q(T_2) < 0$.\\
Having defined the Symanzik polynomials, we can collect some of their elementary properties:
\begin{itemize}
\item The dependence on masses and external momenta is solely given by $\phi_\Gamma$ whereas $\psi_\Gamma$ is independent of physical quantities.
\item Both Symanzik polynomials are homogeneous in the Schwinger parameters. The degree of $\psi_\Gamma$ is $L$ and that of $\phi_\Gamma$ is $L+1$.
\item $\psi_\Gamma$ and $\varphi_\Gamma$ are linear in every single $\alpha_e$. $\phi_\Gamma$ is at most quadratic in the Schwinger parameters (if $m_e \neq 0$).
\item For a product of graphs $\Gamma = \prod_i \gamma_i$ the polynomials $\psi_\Gamma$ and $\varphi_\Gamma$ can be decomposed as follows:
\begin{align}\label{eq:produkt zerlegung symanzik polynome}
\hspace{-2mm}\psi_\Gamma = \prod_i \psi_{\gamma_i}  \mbox{ and }  \varphi_\Gamma = \sum_i \varphi_{\gamma_i} \prod_{\substack{j \\ j\neq i}} \psi_{\gamma_j} .
\end{align}
\end{itemize}
Note that in literature the graph polynomials are also denoted as $\mathcal{U} = \psi_\Gamma$ and $\mathcal{F} = \phi_\Gamma$.

\begin{ex}
As an example, we consider the two-loop graph
\begin{eqnarray} \label{fig:two-loop graph}
\vcenter{\hbox{\includegraphics[scale=0.25]{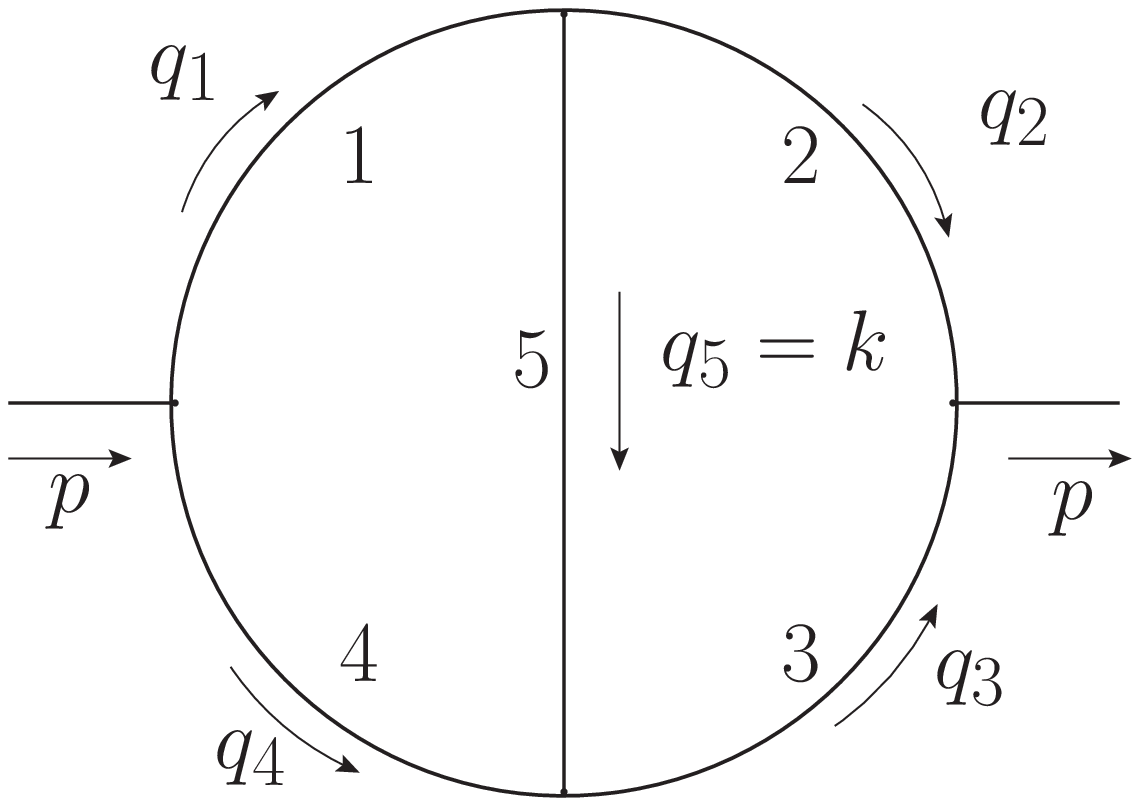}}}
\end{eqnarray}
whose spanning trees and spanning $2$-forests are given in figure \ref{fig:bsp spanning tree} and \ref{fig:bsp spanning 2 forest} respectively.\\
\begin{figure}[H]
	\begin{center}
$
\vcenter{\hbox{\includegraphics[scale=0.4]{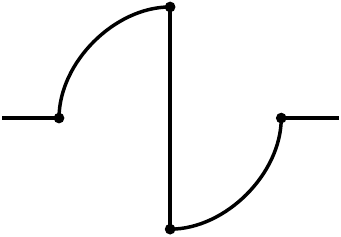}}} \hspace{2mm}
\vcenter{\hbox{\includegraphics[scale=0.4]{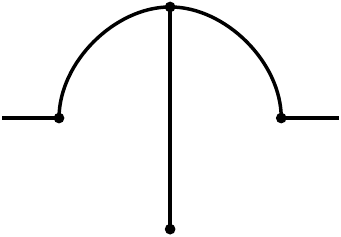}}} \hspace{2mm}
\vcenter{\hbox{\includegraphics[scale=0.4]{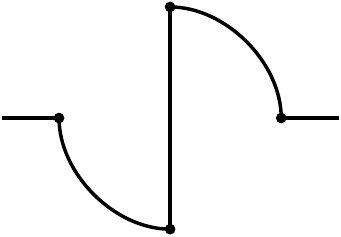}}} \hspace{2mm}
\vcenter{\hbox{\includegraphics[scale=0.4]{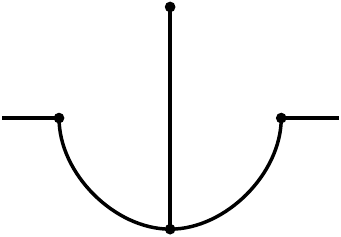}}}
$ \\
\vspace{5mm}
$
\vcenter{\hbox{\includegraphics[scale=0.4]{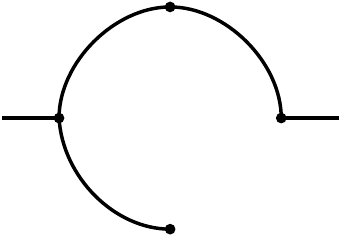}}} \hspace{2mm}
\vcenter{\hbox{\includegraphics[scale=0.4]{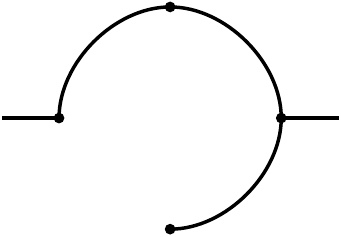}}} \hspace{2mm}
\vcenter{\hbox{\includegraphics[scale=0.4]{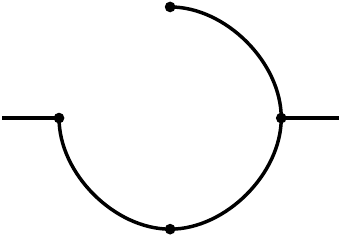}}} \hspace{2mm}
\vcenter{\hbox{\includegraphics[scale=0.4]{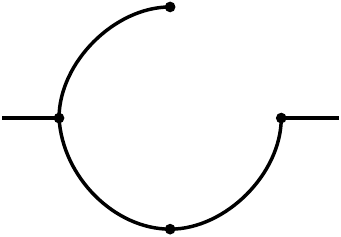}}}
$
	\end{center}
\caption{Set of spanning trees $\mathcal{F}_s^{(1)}$ for the two-loop graph in (\ref{fig:two-loop graph}).}
\label{fig:bsp spanning tree}
\end{figure}
\begin{figure}[H]
	\begin{center}
$
\vcenter{\hbox{\includegraphics[scale=0.35]{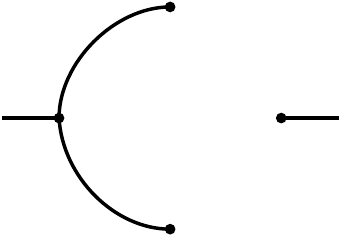}}} \hspace{2mm}
\vcenter{\hbox{\includegraphics[scale=0.35]{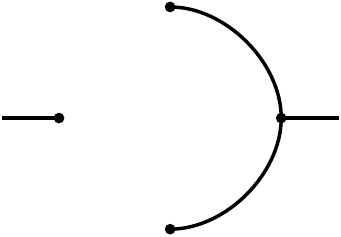}}} \hspace{2mm}
\vcenter{\hbox{\includegraphics[scale=0.35]{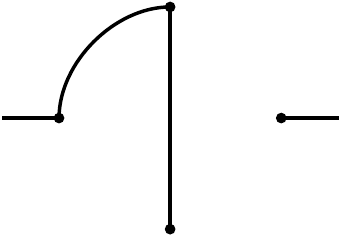}}} \hspace{2mm}
\vcenter{\hbox{\includegraphics[scale=0.35]{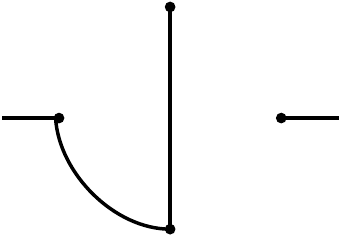}}} \hspace{2mm}
\vcenter{\hbox{\includegraphics[scale=0.35]{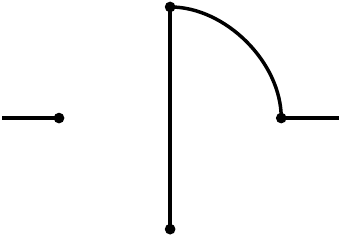}}} 
$ \\
\vspace{5mm}
$
\vcenter{\hbox{\includegraphics[scale=0.35]{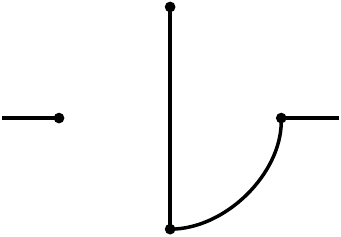}}} \hspace{2mm}
\vcenter{\hbox{\includegraphics[scale=0.35]{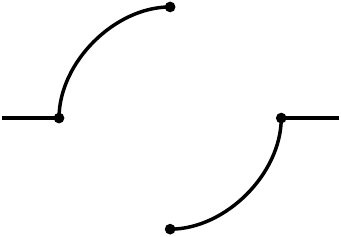}}} \hspace{2mm}
\vcenter{\hbox{\includegraphics[scale=0.35]{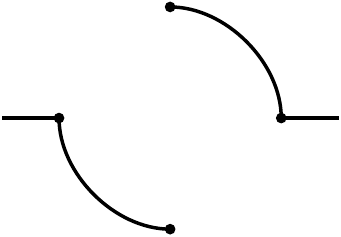}}} \hspace{2mm}
\vcenter{\hbox{\includegraphics[scale=0.35]{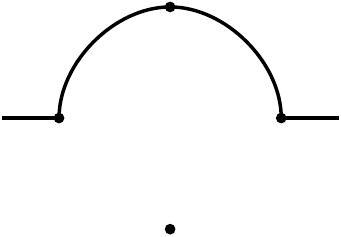}}} \hspace{2mm}
\vcenter{\hbox{\includegraphics[scale=0.35]{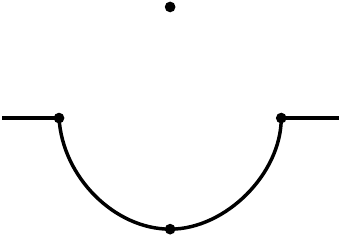}}} 
$
	\end{center}
\caption{Set of spanning $2$-forests $\mathcal{F}_s^{(2)}$ for the two-loop graph in (\ref{fig:two-loop graph}).}
\label{fig:bsp spanning 2 forest}
\end{figure}
The edges are labeled as in the diagram (\ref{fig:two-loop graph}) and carry momentum $q_e$ and mass $m_e$. We impose momentum conservation at each vertex, and the momenta are assumed to flow from left to the right. The first Symanzik polynomial (cf. eq. \ref{eq:first symanzik polynomial}) with respect to the set of spanning trees of the graph (figure \ref{fig:bsp spanning tree}) is given by
\begin{align}
\psi_{\includegraphics[scale=0.1]{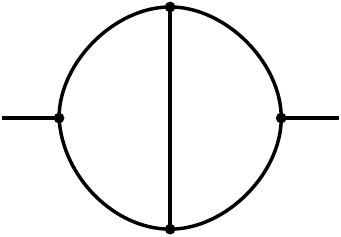}} = (\alpha_1 &+ \alpha_4)(\alpha_2 + \alpha_3) \nonumber\\
&+ \alpha_5 (\alpha_1 + \alpha_2 + \alpha_3 + \alpha_4) .
\end{align}
The single terms in the upper sum are composed of the parameters $\alpha_e$ corresponding to the edges we have to delete to get one of the spanning trees of the graph. Obviously, there are two possible ways to construct a spanning tree out of the graph: Either we delete one edge on the left ($1$ or $4$) and on the right ($2$ or $3$) side, respectively, or we delete the fifth and any of the other edges. The second Symanzik polynomial (defined in eq. (\ref{eq:second symanzik polynomial})) is based on the spanning $2$-forests of the graph given in figure \ref{fig:bsp spanning 2 forest}. There are ten spanning $2$-forests, though only eight of them contribute to $\phi_{\includegraphics[scale=0.1]{bsp/bsp-spanning-tree.pdf}}$. The last two of them do not show up in the polynomial since the sum of the momenta flowing from one tree to the other is zero. Thus,
\begin{align}
\phi_{\includegraphics[scale=0.1]{bsp/bsp-spanning-tree.pdf}} =& \left[(\alpha_1 + \alpha_4) \alpha_2 \alpha_3 + (\alpha_2 + \alpha_3)\alpha_1 \alpha_4 \right. \nonumber\\
&\hspace{1mm}\left.+ (\alpha_1 + \alpha_2)(\alpha_3 + \alpha_4)\alpha_5\right] s \nonumber\\
&\hspace{1mm}+ \psi_{\includegraphics[scale=0.1]{bsp/bsp-spanning-tree.pdf}}  \sum_{e=1}^5 \alpha_e m_e^2
\end{align}
where we used $q_1 + q_4 = -(q_2 + q_3)$ (momentum conservation), and $s = (q_1 + q_4)^2 = (q_2 + q_3)^2$ denotes the center of mass energy. 
\end{ex}

\subsection{The Hopf algebra of rooted trees} \label{subsec:hopf algebra}

The aim of this section is to establish an algebra on the set of Feynman graphs. It was discovered in \cite{Kr:StructurePQFT} that the fundamental mathematical structure on which perturbative renormalization is based on is a Hopf algebra. \\
For the reader not acquainted with the notion of (co-, bi-, Hopf) algebras, some basic definitions are given in appendix \ref{app:algebras}.\\
In section \ref{subsec:graph theoretic foundations} we already introduced the concept of rooted trees and denoted its set by $\mathcal{T}_r$, while $\mathcal{F}_r$ is the set of all rooted forests, i.e. the set of all disjoint unions of rooted trees. The empty tree
\footnote{Note that, by abuse of notation, $\mathbb{I}$ denotes the unit map as well as the empty tree.}
(or empty forest) is denoted by $\mathbb{I} \coloneqq \emptyset$ and has weight zero $|\mathbb{I}| = 0$. We consider a Hopf algebra $H_r$ over $\mathbb{Q}$ generated by the elements of $\mathcal{T}_r$ (including the empty tree $\mathbb{I}$) and define its Hopf algebra structure $\left(H, m, \mathbb{I}, \Delta, \hat{\mathbb{I}}, S\right)$ as follows:
\begin{itemize}
\item For $T_1, T_2 \in \mathcal{T}_r$ the product $m(T_1 \otimes T_2) = T_1 T_2$ is given by the forest $T_1 \cup T_2$, that is the disjoint union of the graphs.
\item The unit map $\mathbb{I}: \mathbb{Q}\rightarrow H_r^{(0)}$ sends $q\in \mathbb{Q}$ to $q\cdot\mathbb{I} \in H_r^{(0)}$.
\item The coproduct on a tree $T \in \mathcal{T}_r$ is defined through 
\begin{align} \label{eq:coproduct rooted tree}
\Delta (T) = &\mathbb{I}\otimes T + T \otimes\mathbb{I} \nonumber\\
&+ \sum_{c \in \mathcal{C}(T)} P^c(T) \otimes R^c(T)
\end{align}
where the sum runs over all admissible cuts $c$ of the tree, whose set $\mathcal{C}$ is given by
\footnote{For reminding the notation, see definition \ref{def:graph}.}
\begin{align}
\mathcal{C}(T) =& \Big\{c\subsetneq E(T): |c \cap (r,v)| \leq 1\nonumber\\
&\hspace{5mm} \forall v \in V(T), c \neq \emptyset\Big\}. 
\end{align}
By making a cut $c$, one or more edges of $T$ are removed and the tree decomposes in a pruned part and a part still containing the root, denoted by $P^c(T)$ and $R^c(T)$, respectively (see example \ref{ex:admissible cuts and antipode} below). For a product of trees, i.e. a forest $f = \cup_i T_i$, we have $\Delta (f) = \prod_i \Delta (T_i)$. The coassociativity of $\Delta$ was shown in \cite{Kr:StructurePQFT}.
\item The counit map $\hat{\mathbb{I}}: H_r \rightarrow \mathbb{Q}$, defined by $\hat{\mathbb{I}}(T) = \begin{cases} 0,\,\, T\neq \mathbb{I} \\ 1,\,\, T = \mathbb{I} \end{cases}$, sends everything that is not the empty tree to zero.
\item A recursive relation for the antipode $S$ acting on a tree can be derived by using $S(\mathbb{I}) = \mathbb{I}$ and equations (\ref{eq:hopf algebra condition}) and (\ref{eq:coproduct rooted tree}), obtaining 
\begin{align}
m\left(S\otimes \operatorname{id}_{H_r}\right) \Delta (T) &= S\left(\mathbb{I}\right) T + \mathbb{I} S(T) \nonumber\\
&\hspace{-10mm}+ \sum_{c \in \mathcal{C}(T)} S\left(P^c(T)\right)R^c(T)\nonumber\\
&= \mathbb{I}\left(\hat{\mathbb{I}}(T)\right) = 0
\end{align}
and thus
\begin{align}
S(T) &= -T - \sum_{c\in \mathcal{C}(T)} S\left(P^c(T)\right)R^c(T)\nonumber\\
& = -T - \sum_{c\in \mathcal{C}(T)} P^c(T) S\left(R^c(T)\right).
\end{align}
For a forest $f$, the antipode is given by $S(f) = S(T_1 \dots T_k) = S(T_k)\dots S(T_1)$.
\end{itemize}
\begin{ex} \label{ex:admissible cuts and antipode}
Take the tree
\begin{eqnarray}
\vcenter{\hbox{\includegraphics[scale=0.3]{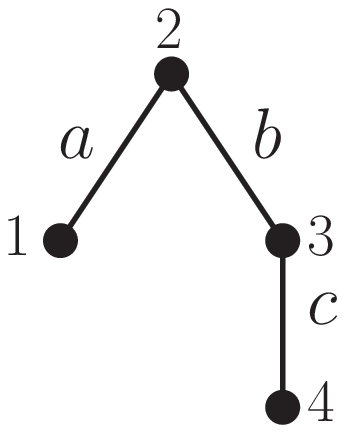}}}
\end{eqnarray}
with edge set $T^{[0]} = \{a, b, c\}$ and vertex set $T^{[1]} = \{1, 2, 3, 4\}$. The set of admissible cuts is given by $\mathcal{C} \left(T\right) = \left\{a, b, c, (a,b), (a, c) \right\}$ or pictorially
\begin{eqnarray} \mathcal{C}\left(T\right) = \left\{
\vcenter{\hbox{\includegraphics[scale=0.2]{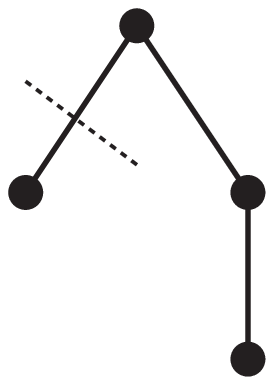}}}, \hspace{2mm}
\vcenter{\hbox{\includegraphics[scale=0.2]{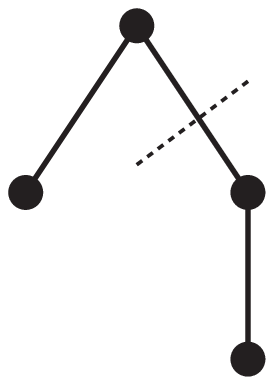}}}, \hspace{2mm}
\vcenter{\hbox{\includegraphics[scale=0.2]{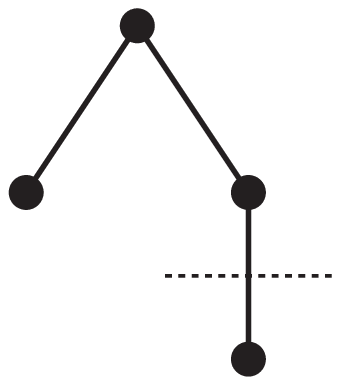}}}, \hspace{2mm}
\vcenter{\hbox{\includegraphics[scale=0.2]{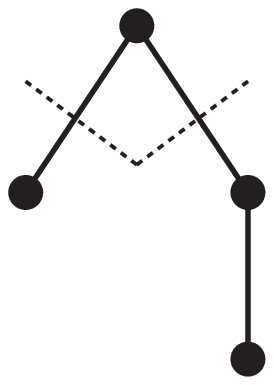}}}, \hspace{2mm}
\vcenter{\hbox{\includegraphics[scale=0.2]{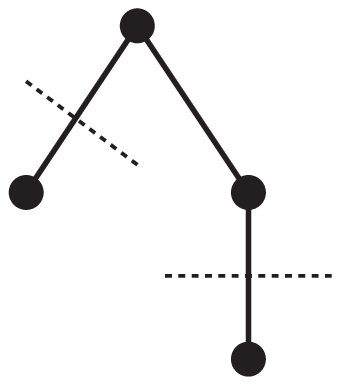}}}
\right\} .
\end{eqnarray}
Therefore, the coproduct of the tree yields
\begin{align}
\Delta\left(\vcenter{\hbox{\includegraphics[scale=0.2]{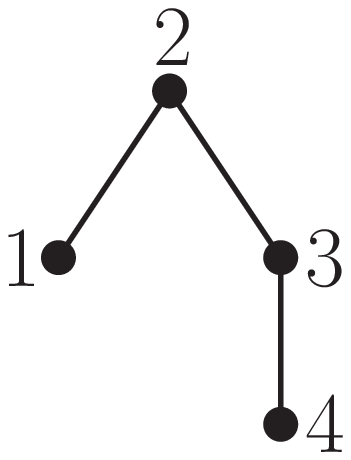}}}\right) =& 
\vcenter{\hbox{\includegraphics[scale=0.2]{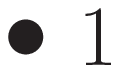}}} \otimes \vcenter{\hbox{\includegraphics[scale=0.2]{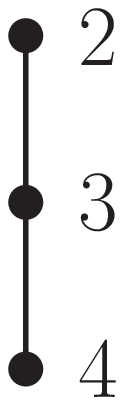}}}  + 
\vcenter{\hbox{\includegraphics[scale=0.2]{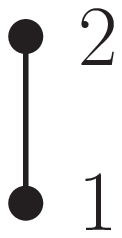}}} \otimes \vcenter{\hbox{\includegraphics[scale=0.2]{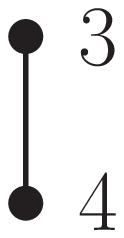}}} \nonumber\\
& \hspace{1mm}+ 
\vcenter{\hbox{\includegraphics[scale=0.2]{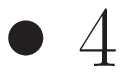}}} \otimes
\vcenter{\hbox{\includegraphics[scale=0.2]{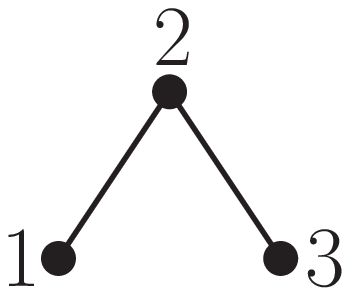}}}  + 
\vcenter{\hbox{\includegraphics[scale=0.2]{bsp/bsp-coproduct-1.eps}}} \vcenter{\hbox{\includegraphics[scale=0.2]{bsp/bsp-coproduct-34.eps}}}  \otimes
\vcenter{\hbox{\includegraphics[scale=0.2]{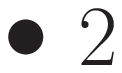}}} \nonumber\\
&\hspace{1mm}+ 
\vcenter{\hbox{\includegraphics[scale=0.2]{bsp/bsp-coproduct-1.eps}}} \vcenter{\hbox{\includegraphics[scale=0.2]{bsp/bsp-coproduct-4.eps}}} \otimes
\vcenter{\hbox{\includegraphics[scale=0.2]{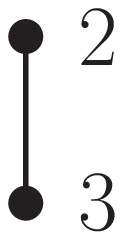}}} ,
\end{align}
and the antipode turns out to be
\begin{align}
S\left(\vcenter{\hbox{\includegraphics[scale=0.2]{bsp/bsp-coproduct-1234.eps}}}\right) = &-
\vcenter{\hbox{\includegraphics[scale=0.2]{bsp/bsp-coproduct-1234.eps}}} + 
\vcenter{\hbox{\includegraphics[scale=0.2]{bsp/bsp-coproduct-1.eps}}}\vcenter{\hbox{\includegraphics[scale=0.2]{bsp/bsp-coproduct-234.eps}}} + 
\vcenter{\hbox{\includegraphics[scale=0.2]{bsp/bsp-coproduct-34.eps}}}\vcenter{\hbox{\includegraphics[scale=0.2]{bsp/bsp-coproduct-12.eps}}}  \nonumber\\
&\hspace{-10mm} - \vcenter{\hbox{\includegraphics[scale=0.2]{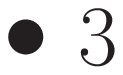}}}\vcenter{\hbox{\includegraphics[scale=0.2]{bsp/bsp-coproduct-4.eps}}}\vcenter{\hbox{\includegraphics[scale=0.2]{bsp/bsp-coproduct-12.eps}}} + 
\vcenter{\hbox{\includegraphics[scale=0.2]{bsp/bsp-coproduct-4.eps}}}\vcenter{\hbox{\includegraphics[scale=0.2]{bsp/bsp-coproduct-123.eps}}}  -
\vcenter{\hbox{\includegraphics[scale=0.2]{bsp/bsp-coproduct-1.eps}}}\vcenter{\hbox{\includegraphics[scale=0.2]{bsp/bsp-coproduct-34.eps}}}\vcenter{\hbox{\includegraphics[scale=0.2]{bsp/bsp-coproduct-2.eps}}} \nonumber\\
&- 
\vcenter{\hbox{\includegraphics[scale=0.2]{bsp/bsp-coproduct-1.eps}}}\vcenter{\hbox{\includegraphics[scale=0.2]{bsp/bsp-coproduct-3.eps}}}\vcenter{\hbox{\includegraphics[scale=0.2]{bsp/bsp-coproduct-4.eps}}}\vcenter{\hbox{\includegraphics[scale=0.2]{bsp/bsp-coproduct-2.eps}}}  - 
\vcenter{\hbox{\includegraphics[scale=0.2]{bsp/bsp-coproduct-1.eps}}}\vcenter{\hbox{\includegraphics[scale=0.2]{bsp/bsp-coproduct-4.eps}}}\vcenter{\hbox{\includegraphics[scale=0.2]{bsp/bsp-coproduct-23.eps}}}
\end{align}
where we used that $S(T_2 T_1) = S(T_1) S(T_2)$,
$S\left(\vcenter{\hbox{\includegraphics[scale=0.3]{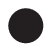}}}\right) = -\vcenter{\hbox{\includegraphics[scale=0.3]{bsp/bsp-coproduct-ladder1.eps}}}$
and
$S\left(\vcenter{\hbox{\includegraphics[scale=0.3]{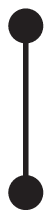}}}\right) = - \vcenter{\hbox{\includegraphics[scale=0.3]{bsp/bsp-coproduct-ladder2.eps}}} + \vcenter{\hbox{\includegraphics[scale=0.3]{bsp/bsp-coproduct-ladder1.eps}}} \vcenter{\hbox{\includegraphics[scale=0.3]{bsp/bsp-coproduct-ladder1.eps}}}$
.
\end{ex}
It was, for example, shown in \cite{Lucia} that the upper definition of $\left(H, m, \mathbb{I}, \Delta, \hat{\mathbb{I}}, S\right)$ gives a commutative, non-cocommutative, connected, and graded Hopf algebra with a natural grading given by the weight (= the node number) of the rooted trees. Taking $|T|$ to be the weight of the rooted tree $T$, the weight of a forest is just $|f| = \sum_i |T_i|$ for $f = \cup_i T_i$. Defining subspaces
\begin{gather}
H_r^{(n)} = \operatorname{span}_\mathbb{Q}\left\{f \in \mathcal{F}_r^{(n)}\right\}
\\
\mbox{ with } \hspace{1mm}\mathcal{F}_r^{(n)} = \left\{f \in \mathcal{F}_r: |f| = n\right\} \hspace{2mm} \forall n\in\mathbb{N}_0 \nonumber
\end{gather}
$H_r$ decomposes as
\begin{align}
H_r = \bigoplus_{n\in\mathbb{N}_0} H_r^{(n)}
\end{align}
which defines a grading on $H_r$. Another subspace of the Hopf algebra of rooted trees is the augmentation ideal
\begin{align}
\operatorname{Aug}_{H_r} \coloneqq \bigoplus_{n \in \mathbb{N}} H_r^{(n)} = \bigoplus_{n = 1}^\infty H_r^{(n)} = \ker \hat{\mathbb{I}}
\end{align}
given by the kernel of the counit. An important endomorphism of $H_r$ is the grafting operator $B_+: H_r \rightarrow \operatorname{span}_\mathbb{Q}(\mathcal{T}_r) \subset H_r$. This operator creates a new root and joins the roots of its arguments to it, returning a single tree:
\begin{gather}
B_+(\mathbb{I}) = \vcenter{\hbox{\includegraphics[scale=0.3]{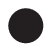}}}
\hspace{2mm}\mbox{ and }\\
B_+(T_1\dots T_n) = \vcenter{\hbox{\includegraphics[scale=0.3]{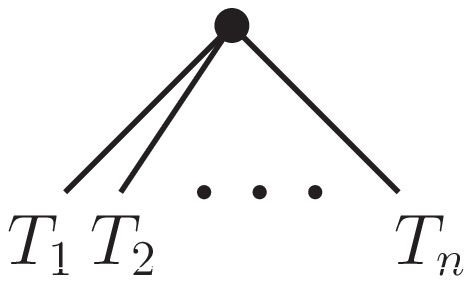}}}.
\end{gather}
The operator $B_+$ satisfies the relation
\begin{align} \label{eq:coproduct grafting operator}
\Delta \circ B_+ = B_+ \otimes \mathbb{I} + (\operatorname{id} \otimes B_+) \circ \Delta
\end{align}
which can be regarded as a recursive definition of the coproduct since every tree can be written as $T = B_+(X)$ and $\Delta(\mathbb{I}) = \mathbb{I} \otimes \mathbb{I}$.
\begin{rem}
In fact, equation (\ref{eq:coproduct grafting operator}) implies that $B_+$ is a $1$-cocycle in the Hochschild cohomology of $H_r$, see \cite{CoKr:Renorm98} or \cite{Foissy:IntroHopf} for example.
\end{rem}
At the end of this section, we want to introduce a sub-Hopf algebra of $H_r$, namely the Hopf algebra of ladders, we will come back to, later. Generally, a sub-Hopf algebra of a graded Hopf algebra $\left(H, m, \mathbb{I}, \Delta, \hat{\mathbb{I}}, S\right)$ with $H = \oplus_i H^{(i)}$ is defined as the subspace $\tilde{H} \subset H$, such that $\tilde{H}$ has a Hopf algebra structure $\left(\tilde{H}, m, \mathbb{I}, \Delta, \hat{\mathbb{I}}, S\right)$ and a grading $\tilde{H} = \oplus_i \left(\tilde{H}\cap H^{(i)}\right)$. We define a ladder of weight $k$ by $\lambda_k \coloneqq (B_+)^k(\mathbb{I})$, which is the $k$-fold application of the grafting operator on the empty tree. Thus, ladders can be generated iteratively through $\lambda_k = B_+ (\lambda_{k-1})$ with $\lambda_0 \coloneqq \mathbb{I}$. Therefore, the diagrams take the form:
\begin{align*}
\lambda_0 = \mathbb{I}, \hspace{2mm} 
\lambda_1 = \vcenter{\hbox{\includegraphics[scale=0.2]{graphs/ladder-1.eps}}}, \hspace{2mm}
\lambda_2 = \vcenter{\hbox{\includegraphics[scale=0.2]{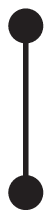}}}, \hspace{2mm}
\dots, \hspace{2mm}
\lambda_k = \left.\vcenter{\hbox{\includegraphics[scale=0.2]{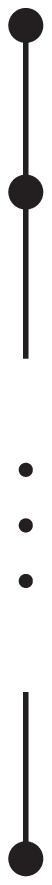}}} \right\} k\mbox{-times} .
\end{align*}
The sub-Hopf algebra $H_L$, generated by the ladders, decomposes in the subspaces $H_L^{(n)} \subset H_r^{(n)}$ which consist of the elements of weight $n$. The coproduct on $H_L$ is given by $\Delta(\lambda_k) = \sum_{j=0}^k \lambda_j \otimes \lambda_{k-j}$.

\section{Parametric renormalization}\label{sec:parametric renorm}
%
We now let $\Gamma$ be a Feynman graph in a scalar quantum field theory with arbitrary oriented edges. 
Generally, the Feynman rules of the underlying theory assign an integral to the graph whose parametric representation is given by \cite{Panzer:PhD}
\begin{align}\label{eq:parametric rep}
\Phi\left(\Gamma\right) = \prod_{e\in\Gamma_{\mbox{\tiny int}}{[1]}} \int_{\mathbb{R}_+} \frac{\alpha_e^{a_e -1} \operatorname{d}\alpha_e}{\Gamma(a_e)}\cdot\frac{e^{-\phi_\Gamma/\psi_\Gamma}}{\psi_\Gamma^{D/2}}.
\end{align}
and which can be obtained from a Schwinger parametrization of $\Gamma$. The exponents $a_e$ are assumed to be equal to $1$. 
Thus, the parametric integral becomes a function of the Schwinger parameters $\alpha_e$, the squared particle masses $m_e^2$, and the scalar products of the momenta $p_i \cdot p_j$ ($i, j \in \Gamma$). Following \cite{BroKr:AnglesScales}, we want to rescale the Feynman rules $\Phi\left(\Gamma\right)$ by a parameter $S$. Therefore, we introduce dimensionless scattering angles $\left\{\Theta\right\} = \left\{\Theta_{ij}, \Theta_e\right\}$ given by the scaled variables
\begin{align} \label{eq:scattering angles}
\Theta_{ij} = \frac{p_i \cdot p_j}{S} \hspace{5mm} \mbox{ and } \hspace{5mm} \Theta_e = \frac{m_e^2}{S} .
\end{align}
The variable $S$ sets the scale of the graph $\Gamma$ defined by
\begin{align} \label{eq:scale}
S \coloneqq \sum_{e\in\Gamma_{\mbox{\tiny ext}}^{[1]}} p_e^2 ,
\end{align}
such that $S > 0$ and $S = 0$ only if all external momenta collectively vanish
\footnote{Here and throughout this article we explicitly exclude lightlike particles.}
. The rescaled Feynman rules then can be written as a function of the scale variable and the angles
\begin{align}
\Phi\left(\Gamma\right) \left\{S\Theta_{ij}, S\Theta_e \right\} \rightarrow \Phi\left(\Gamma\right) \left\{S, \Theta_{ij}, \Theta_e \right\} 
\end{align}
and the integral evaluates to 
\begin{align} \label{eq:rescaled integral}
\Phi\left(\Gamma\right)\left\{S, \Theta\right\} = \prod_{e\in\Gamma_{\mbox{\tiny int}}^{[1]}} \int_{\mathbb{R}_+} \operatorname{d}\alpha_e \frac{e^{-S\frac{\phi_\Gamma(\Theta)}{\psi_\Gamma}}}{\psi_\Gamma^{D/2}} 
\end{align}
with
\begin{align}
\phi_\Gamma\left(\Theta\right) &= \frac{\varphi_\Gamma}{S} + \psi_\Gamma \sum_{e\in\Gamma_{\mbox{\tiny int}}^{[1]}} \alpha_e \frac{m_e^2}{S} \nonumber \\ 
&= \varphi_\Gamma\left(\Theta\right) + \psi_\Gamma \sum_{e\in\Gamma_{\mbox{\tiny int}}^{[1]}} \alpha_e \Theta_e .
\end{align}
Since $\psi_\Gamma$ is independent of physical quantities it is not affected by the rescaling.\\
To carry out one of the integrations we insert $1 = \int_0^\infty \operatorname{d}t \, \,\delta\left(t - \sum_e \lambda_e \alpha_e\right)$ with $\lambda_e \geq 0$ not all zero into (\ref{eq:rescaled integral}) and substitute $\alpha \to t\alpha_e$ which leads to $\prod_{e\in\Gamma_{\mbox{\tiny int}}^{[1]}} \operatorname{d}\alpha_e \rightarrow t^{E_\Gamma - 1} \operatorname{d}t \wedge \Omega_\Gamma$ where $\wedge$ denotes the exterior product (or wedge product) and the ($E_\Gamma-1$)-form $\Omega_\Gamma$ defines the volume form $\Omega_\Gamma \coloneqq \sum_{i = 1}^{E_\Gamma} \left(-1\right)^{i+1} \alpha_i \operatorname{d}\alpha_1\wedge\dots\wedge\widehat{\operatorname{d}\alpha_i} \wedge\dots\wedge\operatorname{d}\alpha_{E_\Gamma}$ in projective space $\mathbb{P}_\Gamma \coloneqq \mathbb{P}^{E_\Gamma - 1}\left(\mathbb{R}_+\right)$ (cf. \cite{BroKr:AnglesScales} and \cite{Panzer:PhD}). The circumflex accent $\widehat{\hspace{5mm}}$ means that the argument is omitted. The projective integral finally takes the form
\begin{align} \label{eq:projective rep}
\Phi\left(\Gamma\right)\left\{S, \Theta\right\} = \int_{\mathbb{R}_+} \int_{\mathbb{P}_\Gamma} \frac{\operatorname{d}t}{t} \wedge \frac{e^{-tS\frac{\phi_\Gamma(\Theta)}{\psi_\Gamma}}}{t^{\omega_D/2} \psi_\Gamma^{D/2}} \Omega_\Gamma
\end{align}
with $\omega_D$ the superficial degree of divergence
\footnote{In $\phi^k_D$-theory, the superficial degree of divergence is given by $\omega_D = 2\cdot E_\Gamma - D\cdot L$ since the weight of the edges and vertices is $\omega(e) = 2$ and $\omega(v) = 0$, respectively. 
}.
Now we want to perform the $t$-integration to get rid of the exponential. Therefore, we have to distinguish between the case $\omega_D > 0$, where the integral converges, and the case of ultraviolet divergence, i.e. $\omega_D \leq 0$. We are only interested in the latter case, thus we have to renormalize the integral. We apply kinetic renormalization conditions to $\Phi\left(\Gamma\right)$, that is to say that the renormalized amplitude of the graph $\Gamma$ vanishes at a chosen reference or renormalization point $\left\{S_0, \Theta_0\right\}$, as well as all of its first $\omega_D$ derivatives in the Taylor expansion around that point.\\
In the logarithmic divergent case ($\omega_D = 0$), this condition can be implemented by modifying $\Phi\left(\Gamma\right)$ as follows
\begin{align} 
\Phi\left(\Gamma\right)\left\{S, \Theta\right\} \rightarrow &\big[\Phi\left(\Gamma\right)\left\{S, \Theta\right\} \nonumber\\
&\hspace{10mm }- \Phi\left(\Gamma\right)\left\{S_0, \Theta_0\right\}\big] .
\end{align}
Hence, the overall divergence can be cured by a subtraction at the reference point.\\
As it was shown in \cite{BroKr:AnglesScales}, the renormalized integral decomposes into angle- and scale-dependent parts. In particular, the renormalized Feynman rules can be written as a polynomial in the scaling parameter $L = \ln (S/S_0)$
\begin{align}
\Phi_R \left(\Gamma\right) = \sum_{j=0}^{\operatorname{cor}\left(\Gamma\right)} c_j^\Gamma\left(\Theta, \Theta_0\right) L^j .
\end{align}
The integer $\operatorname{cor}\left(\Gamma\right)$ is called the co-radical degree of $\Gamma$ defined as the maximal integer $j_{\operatorname{max}}\in \mathbb{N}$, such that
\begin{align}
\tilde\Delta^{j_{\operatorname{max}} - 1} \Gamma \neq 0 \hspace{2mm}\mbox{ and }\hspace{2mm} \tilde\Delta^j \Gamma = 0,\hspace{2mm} \forall j\geq j_{\operatorname{max}}
\end{align} 
in which $\tilde{\Delta}^j$ is the iterated reduced coproduct (see equations (\ref{eq:reduced coproduct}) and (\ref{eq:iterated rdeuced coproduct})). Therefore, $\operatorname{cor}\left(\Gamma\right)$ equals the weight $\left|T\left(\Gamma\right)\right|$ of the rooted tree $T$ associated with the subgraph-structure of the graph $\Gamma$ and in principle indicates how many divergent subgraphs are nested in $\Gamma$.\\
The coefficients $c_j^\Gamma$ can be determined by calculating the renormalized Feynman integral. In the case of Feynman diagrams with more than one independent loop we have to proceed recursively in order to eliminate all possible subdivergences. Of course, this recursion rapidly becomes more complicated. But in the case of momentum subtraction schemes there is an elegant solution to the recursion problem provided by Zimmermann, the so-called forest formula. Applying this formula to the graph $\Gamma$ and expanding the resulting integral as a power series in the scaling parameter $L$ allows us to determine the coefficients $c_j^\Gamma$. \\
Assuming that $\Gamma$ has not only logarithmic subdivergences but is additionally overall divergent, the forest formula yields \cite{Sars:PhD}
\begin{align} \label{eq:renorm feynman rules}
&\Phi_R\left(\Gamma\right)\left\{S, S_0, \Theta, \Theta_0\right\} \nonumber\\ 
&= \sum_{f \in \mathcal{F}\left(\Gamma\right)} (-1)^{\# f} \Big[ \Phi\left(f\right)\left\{S_0, \Theta_0\right\}\Phi\left(\Gamma/f\right)\left\{S, \Theta\right\} \nonumber \\
&\hspace{12mm}- \Phi\left(f\right)\left\{S_0, \Theta_0\right\}\Phi\left(\Gamma/f\right)\left\{S_0, \Theta_0\right\}\Big]
\end{align}
for the renormalized integrand, where the sum is over all forests $f$ of the graph $\Gamma$, also including the empty
\footnote{Note that for the empty forest $f = \left\{\emptyset\right\}$, the graph polynomials are defined as $\psi_\emptyset = 1$ and $\phi_\emptyset(\Theta) = 0$.}
 one but excluding the forest containing $\Gamma$ itself, and  $\# f$ denotes the number of connected components of $f$.\\
If we plug (\ref{eq:projective rep}) in our formula (\ref{eq:renorm feynman rules}), we see that the resulting expression is singular at $t = 0$ due to the $\frac{\operatorname{d} t}{t}$-integration over $\mathbb{R}_+$. Nevertheless, it can be regularized by introducing a regulator $c$, using that, for sufficiently small $c>0$,
\begin{align}
\hspace{-2mm}\int_c^\infty \hspace{-1mm} \frac{e^{-tX}}{t} \operatorname{d}t = - \ln c - \ln X - \gamma_E + \mathcal{O}(c\ln c)
\end{align}
with $X>0$ fixed and $\gamma_E$ the Euler-Mascheroni constant.
 To take the limit $c \to 0$, we have to subtract the integral at $X_0$ which yields
\begin{align}
\lim\limits_{c\to 0}{ \int_c^\infty \left[e^{-tX} - ^{-tX_0}\right] \frac{\operatorname{d}t}{t} = - \ln (X/X_0)} .
\end{align}
Therefore, we found a way to carry out the $t$-integration. Applying the formula above to the integral (\ref{eq:renorm feynman rules}) with (\ref{eq:projective rep}) plugged in finally delivers 
\begin{align} \label{eq:renorm fr projective space}
\Phi_R\left(\Gamma\right) =& - \int_{\mathbb{P}_\Gamma} \sum_{f\in\mathcal{F}\left(\Gamma\right)} \Bigg[(-1)^{\# f} \frac{1}{\psi_{\Gamma/f}^{D/2} \psi_f^{D/2}} \nonumber\\
&\hspace{-5mm} \times\ln\left(\frac{S\phi_{\Gamma/f}\psi_f + S_0 \phi_f^0\psi_{\Gamma/f}}{S_0\phi_{\Gamma/f}^0\psi_f + S_0 \phi_f^0\psi_{\Gamma/f}}\right)\Bigg]\Omega_\Gamma
\end{align}
for the renormalized Feynman rules in projective form.

\section{The $L$-linear term of signed graph permuations}\label{sec:L-linear term}

We are interested in the coefficient $c_1^\Gamma (\Theta, \Theta_0)$ of the term of $\Phi_R$ linear in $L$. Therefore, we first have to differentiate (\ref{eq:renorm fr projective space}) with respect to $L$ at $S = S_0$ (or equivalent $L=0$)
\begin{align}
c_1^\Gamma (\Theta, \Theta_0) = \left.\frac{\partial \Phi_R\left(\Gamma\right)}{\partial L}\right|_{L = 0}  \hspace{-3mm} = \left.S\frac{\partial \Phi_R\left(\Gamma\right) }{\partial S} \right|_{S = S_0}
\end{align}
which yields 
\begin{align}\label{eq:linear renormalized feynman rules}
&\Phi_R^{(1)}\left(\Gamma\right) = c_1^\Gamma (\Theta, \Theta_0)= \nonumber\\
& \int_{\mathbb{P}_\Gamma} \sum_{f \in\mathcal{F} \left(\Gamma\right)} \Bigg[(-1)^{\# f} \frac{1}{\psi_{\Gamma/f}^2 \psi_f^2}  \nonumber\\
&\hspace{25mm}\times \frac{\phi_{\Gamma/f}\psi_f}{\phi_{\Gamma/f}\psi_f + \phi_f \psi_{\Gamma/f}}\Bigg]\Omega_\Gamma .
\end{align}
Without loss of generality we set the dimension of spacetime equal to four, $D = 4$. 
If we assume that $L$ is very small, this term gives us the main contribution to $\Phi_R$ together with the $L$-independent term. In \cite{BroKr:AnglesScales}, the upper term was also derived and discussed in great detail.\\

For symmetric sums over permutations of graph insertions (symmetric flags) it was already shown in \cite{KrKr:FiltrationsDSE} that the angle-dependence drops out in the $L$-linear term of the renormalized Feynman rules, and we will prove that this is also true in the case of antisymmetric flags. Moreover, we present a formula which allows us to compute all (angle-independent) terms surviving in the sum. Thereby, the problem of finding all forests of a graph is boiled down to the much more simple task of figuring out all possible decomposition of the co-radical degree of the graph into positive integers.\\
Analogous to \cite{KrKr:FiltrationsDSE} we, therefore, define:
\begin{defi}[Flag]
A Hopf algebra element $\Gamma$ of co-radical degree $\operatorname{cor}(\Gamma) = r_\Gamma$ is said to be a flag if there exists a sequence of primitive graphs $\gamma_i$ with $1\leq i \leq r_\Gamma$ such that
\begin{align}
\tilde{\Delta}^{r_\Gamma - 1} \Gamma = \gamma_1 \otimes \dots \otimes \gamma_{r_\Gamma}.
\end{align}
\end{defi}
If $\Gamma$ is a flag, the corresponding rooted tree is given by the (decorated) ladder
\begin{align}
\lambda_{r_\Gamma}^{(r_\Gamma, \dots , 1)} = \vcenter{\hbox{\includegraphics[scale=0.2]{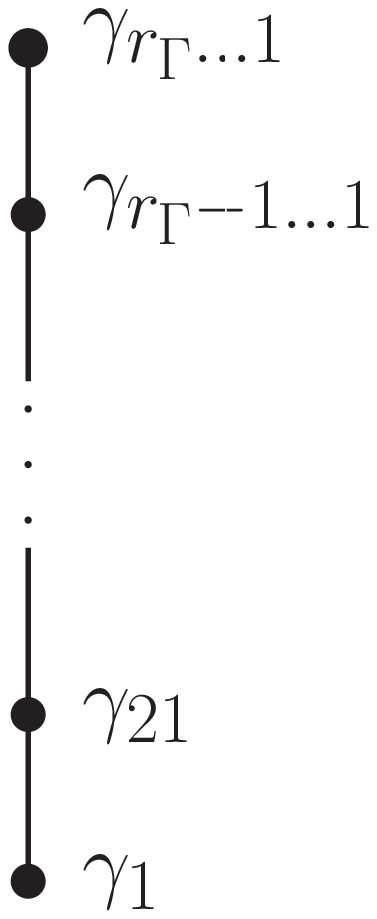}}}
\end{align}
of weight $r_\Gamma$. The expression $\gamma_{r_\Gamma \dots 1}$ is shorthand for the successive nested insertion $\gamma_{r_\Gamma} \leftarrow \left( \dots \leftarrow  \gamma_1 \right)$, meaning that we start with $\gamma_1$, insert it into $\gamma_2$, insert the resulting graph $\gamma_{21}$ into $\gamma_3$, and so on until we end up with inserting $\gamma_{r_\Gamma-1 \dots 1}$ into $\gamma_{r_\Gamma}$ receiving the graph $\Gamma = \gamma_{r_\Gamma \dots 1}$.
In the following it will prove beneficial to label the vertices of the ladder only by the leading index of the subgraph associated with it, i. e.
\begin{align}
\vcenter{\hbox{\includegraphics[scale=0.2]{graphs/flag.eps}}}  \Longleftrightarrow \hspace{7mm}
\vcenter{\hbox{\includegraphics[scale=0.2]{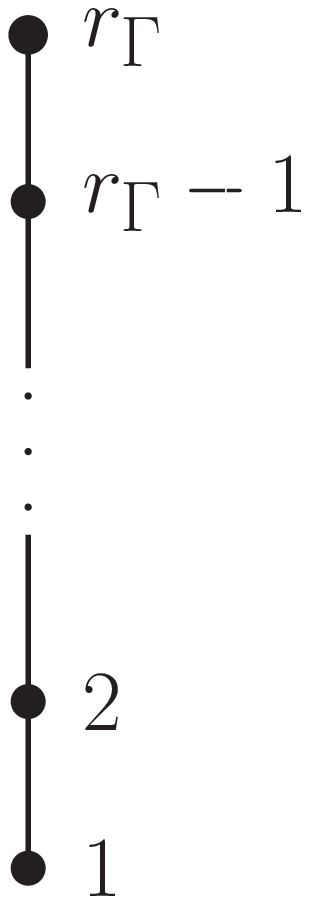}}} .
\end{align}
Let $\Lambda_{r_\Gamma}$ be a sum of $r_\Gamma$ flags $\Lambda^{(i)}$
\begin{align} \label{eq:flag}
\Lambda_{r_\Gamma} = \sum_{i = 1}^{r_\Gamma} \Lambda^{(i)} .
\end{align}
This sum is called a symmetric flag $\Lambda_{r_\Gamma}^+$ if
\begin{align} \label{eq:symmetric flag}
\tilde{\Delta}^{r_\Gamma - 1} \Lambda_{r_\Gamma}^+ = \sum_{\sigma} \gamma_{\sigma (1)} \otimes \dots \otimes \gamma_{\sigma (r_\Gamma)}
\end{align}
where the sum is over all $r_\Gamma !$ permutations of the primitive graphs $\gamma_i$. 
Accordingly, we say that $\Lambda_{r_\Gamma}^-$ is an antisymmetric flag if the $r_\Gamma - 1$-fold application of the reduced coproduct gives
\begin{align}\label{eq:antisymmetric flag}
\tilde{\Delta}^{r_\Gamma - 1} \Lambda_{r_\Gamma}^- &= \sum_{\sigma \mbox{\tiny cycl.}}\sum_{i_1, \dots , i_{r_\Gamma - 1} = 1}^{r_\Gamma-1} \hspace{-2mm} \varepsilon_{i_1 \dots  i_{r_\Gamma -1}} \gamma_{\sigma(i_1)} \otimes \nonumber\\
&\hspace{15mm}\dots \otimes \gamma_{\sigma(i_{r_\Gamma - 1 })} \otimes \gamma_{\sigma(r_\Gamma)} 
\end{align}
for a sequence of primitive graphs $\gamma_i$ with $1\leq i \leq r_\Gamma$ where the sum is over all permutations within the group $\left( \mbox{signed } S_{r_\Gamma-1}\right) \times \left( \mbox{cyclic } S_{r_\Gamma}\right)$. \\
By analogy with the Levi-Cevita-symbol we introduce the tensor
\begin{small}
 \begin{align} \label{eq:epsilon tilde}
\tilde{\varepsilon}_{i_1, \dots , i_{r_\Gamma}} = 
	\begin{cases}
	+1 \mbox{ if } (i_1, \dots , i_{r_\Gamma}) \mbox{ is a cyclic permutation } \\ 
	\hspace{5mm}\mbox{ of } (1, k_2, \dots , k_{r_\Gamma}) \mbox{ where } (k_2, \dots , k_{r_\Gamma })  \\
	\hspace{5mm}\mbox{ is an even permutation of } (2, \dots, r_\Gamma) \mbox{ , } \\
	-1 \mbox{ if } (i_1, \dots , i_{r_\Gamma}) \mbox{ is a cyclic permutation } \\ 
	\hspace{5mm}\mbox{ of } (1, k_2, \dots , k_{r_\Gamma}) \mbox{ where } (k_2, \dots , k_{r_\Gamma })  \\
	\hspace{5mm}\mbox{ is an odd permutation of } (2, \dots, r_\Gamma) \mbox{ , } \\
	\hspace{2.75mm} 0 \mbox{ if at least two indices of } \\
	\hspace{5mm} \mbox{ the set} (i_1, \dots , i_{r_\Gamma}) \mbox{ are equal }
	\end{cases}
\end{align}
\end{small}
defined from the sign of a permutation $\sigma \in S^{\mbox{\tiny signed}}_{r_\Gamma - 1} \times S^{\mbox{\tiny cyclic}}_{r_\Gamma}$. The chosen permutation group ensures that cyclic permutations of the indices $(i_1,\dots , i_{r_\Gamma})$ conserve the sign, as can be seen from the tensor $\tilde{\varepsilon}$ defined above. For $r_\Gamma = \mbox{ odd }$ we need an even number of transpositions to perform a cyclic permutation and therefore $\sigma \in S^{\mbox{\tiny cyclic}}_{r_\Gamma = \mbox{\tiny odd}}$ preserves the sign. Thus 
\begin{align}
S^{\mbox{\tiny signed}}_{r_\Gamma - 1} \times S^{\mbox{\tiny cyclic}}_{r_\Gamma} \Big |_{r_\Gamma = \mbox{\tiny odd}} &= S^{\mbox{\tiny signed}}_{r_\Gamma}  \mbox{ and } \nonumber\\
\tilde{\varepsilon}_{i_1, \dots , i_{r_\Gamma}} \Big |_{r_\Gamma = \mbox{\tiny odd }} &= \varepsilon_{i_1, \dots , i_{r_\Gamma}} .
\end{align}
In terms of ladders, $\Lambda_{r_\Gamma}^-$ can then be written as
\begin{align}
\Lambda_{r_\Gamma}^- &= 
\sum_{i_1, \dots , i_{r_\Gamma} = 1}^{r_\Gamma} \tilde{\varepsilon}_{i_1 \dots  i_{r_\Gamma }} \lambda_{r_\Gamma}^{(i_{r_\Gamma}, \dots , i_1)} \nonumber\\
&= \sum_{i_1, \dots , i_{r_\Gamma} = 1}^{r_\Gamma} \tilde{\varepsilon}_{i_1 \dots  i_{r_\Gamma }}  \vcenter{\hbox{\includegraphics[scale=0.25]{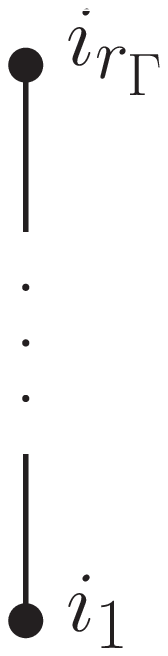}}}
\end{align}
and analogously $\Lambda_{r_\Gamma}^+$.\\
It was already shown in \cite{KrKr:FiltrationsDSE} that the coefficient $\Phi_R^{(1)} \left(\Lambda_{r_\Gamma}^+\right)$ is angle-independent if the renormalization point preserves scattering angles, i.e. $\Theta \equiv \Theta_0$. What we want to assert is:
\begin{prop}
Let $\Lambda_{r_\Gamma}^-$ be a antisymmetric flag as defined in (\ref{eq:antisymmetric flag}). Then, the $L$-linear term $\Phi_R^{(1)} \left (\Lambda_{r_\Gamma}^- \right)$ of the renormalized Feynman rules is independent of the second Symanzik polynomial and, therefore, angle-independent under the assumption that the renormalization point preserves scattering angles.
\end{prop}
In order to prove the upper proposition, we first want to do an explicit example. \\
We are looking at the graph \\ $\gamma_{ijk} = \gamma_i \leftarrow \underbrace{(\gamma_j \leftarrow \gamma_k)}_{\gamma_{jk}}$ of co-radical $r_\Gamma = 3$ with $1 \leq i, j, k \leq 3$ and $i \neq j \neq k$. The decorated rooted tree associated to the graph is given by $\vcenter{\hbox{\includegraphics[scale=0.3]{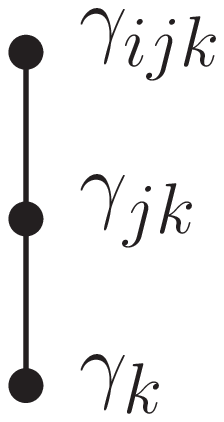}}}$, and the forest-set deduced from it turns out to be
\begin{align} \label{eq:forest N=3}
\mathcal{F}(\gamma_{ijk}) = \left\{\emptyset, \gamma_{jk}, \gamma_k, \gamma_{jk} \cup \gamma_k\right\} .
\end{align}
The coefficient $\Phi_R^{(1)}$ of the $L$-linear term is (see equation (\ref{eq:linear renormalized feynman rules}))
\begin{align}
\Phi_R^{(1)} \left(\gamma_{ijk}\right) &= \int_{\mathbb{P}_\gamma} \Omega_\gamma \cdot I\left(\gamma_{ijk}\right) \hspace{5mm} \mbox{ with } \label{eq:linear renorm fr N=3}\\
I\left(\gamma_{ijk}\right) &= \frac{1}{\psi_{ijk}^2} - \frac{1}{\psi_{ij}^2 \psi_k^2} \frac{\phi_{ij} \psi_k}{\phi_{ij}\psi_k + \phi_k \psi_{ij}} \nonumber\\
& \hspace{1mm}- \frac{1}{\psi_i^2 \psi_{jk}^2} \frac{\phi_i \psi_{jk}}{\phi_i \psi_{jk} + \phi_{jk} \psi_i} \label{eq:integrand N=3}\\
&\hspace{1mm}+ \frac{1}{\psi_i^2 \psi_j^2 \psi_k^2} \frac{\phi_i \psi_j \psi_k}{\phi_i \psi_j \psi_k + \left(\phi_j \psi_k + \phi_k \psi_j\right)\psi_i} \nonumber 
\end{align}
if we assume that the renormalization point preserves scattering angles. To make it more compact, we used the shorthand notation $\psi_{i_1 \dots i_n} \equiv \psi (\gamma_{i_1 \dots i_n})$ and analogue for $\phi$. In the last term of $I\left(\gamma_{ijk}\right)$, which corresponds to the forest $\gamma_{jk}\cup\gamma_k$, we used the decomposition rules (\ref{eq:produkt zerlegung symanzik polynome}) for products of graphs to rewrite the Symanzik polynomials. Now we intend to show the angle-independence of the $L$-linear term for a antisymmetric flag of co-radical degree three. For that purpose we use $\tilde{\varepsilon}_{ijk} = \varepsilon_{ijk} = \frac{1}{3} \left\{\varepsilon_{ijk} + \varepsilon_{jki} + \varepsilon_{kij}\right\}$ to get
\begin{align} \label{eq:antisymmetric integrand N=3}
\varepsilon_{ijk} I &\left(\gamma_{ijk}\right) = \frac{1}{3} \varepsilon_{ijk} \left\{\frac{1}{\psi_{ijk}^2} + \frac{1}{\psi_{jki}^2} + \frac{1}{\psi_{kij}^2}\right. \nonumber\\
&\hspace{-4.5mm}\left. - \frac{1}{\psi_{ij}^2 \psi_k^2} - \frac{1}{\psi_{jk}^2 \psi_i^2} - \frac{1}{\psi_{ki}^2 \psi_j^2} + \frac{1}{\psi_i^2 \psi_j^2 \psi_k^2}\right\} .
\end{align}
As one can see, all $\phi$-dependent terms summed up to unity such that the whole expression is independent of the scattering angles. The first and the second three terms in the upper equation are just cyclic permutations of each other. Therefore, they add up to one term if we sum over all indices $i$, $j$, and $k$. The last term in equation (\ref{eq:antisymmetric integrand N=3}) cancels in the sum because of the sign change due to the Levi-Cevita-tensor.
As a consequence, the antisymmetric sum of the $L$-linear terms of all graphs $\gamma_{ijk}$ yields
\begin{align} \label{eq:result n=3}
\Phi_R^{(1)} \left(\Lambda_3^-\right) &= \sum_{i, j, k = 1}^3 \varepsilon_{ijk} \Phi_R^{(1)} \left(\gamma_{ijk}\right) \nonumber\\
& \hspace{-8mm}= \int_{\mathbb{P}_\gamma} \Omega_\gamma \sum_{i, j, k = 1}^3 \varepsilon_{ijk} \left\{\frac{1}{\psi_{ijk}^2} - \frac{1}{\psi_{ij}^2 \psi_k^2}\right\} .
\end{align}
This expression only depends on the first Symanzik polynomial and is thus independent of the renormalization point.\\
Since we want to prove that this is always the case for antisymmetric flags, let us do this more general. We consider the graph $\gamma_{r_\Gamma \dots 1}$ of co-radical degree $r_\Gamma$ and pick a general forest of it, namely  $f = \left\{\gamma_{i_j \dots i_1} \cup \gamma_{i_k \dots i_1} \cup \gamma_{i_l \dots i_1} \cup \gamma_{i_m \dots i_1} \right\}$ with $r_\Gamma > j > k > l > m > 1$ that generates the term 
\begin{gather} \label{eq:general term}
\frac{\phi_{i_{r_\gamma} \dots i_{j+1}} \psi_{i_j \dots i_{k+1}} \psi_{i_k \dots i_{l+1}} \psi_{i_l \dots i_{m+1}} \psi_{i_m \dots i_{1}} }{\big(\prod_{d \in D} \psi_d^2\big) \times \big(\sum_{d \in D} \phi_d \prod_{\substack{d' \in D, \\  d' \neq d}} \psi_{d'}\big)} \\
\mbox{ with } D = \big\{i_{r_\gamma} \dots i_{j+1}, i_j \dots i_{k+1}, i_k \dots i_{l+1}, \nonumber\\
\hspace{45mm} i_l \dots i_{m+1}, i_m \dots i_{1} \big\} \nonumber
\end{gather}
in $\Phi_R^{(1)}$. Now we have to distinguish between two cases:
\subsubsection*{(1) There are at least two $\psi_{i_s \dots i_{t+1}}$ with $s = t+1$.}\label{vanishing condition}
This means that we have more than one $\psi$ with only one index. Without loss of generality we assume that $k = l+1$ and $l = m+1$. Therefore, the numerator in (\ref{eq:general term}) takes the form
\begin{align}
\phi_{i_{r_\gamma} \dots i_{j+1}} \psi_{i_j \dots i_{k+1}} \psi_{i_k} \psi_{i_l} \psi_{i_m \dots i_{1}} .
\end{align}
When changing $i_k$ and $i_l$, we get a sign flip from the $\tilde{\varepsilon}$-tensor but the whole fraction stays invariant under that permutation and thus these terms cancel in the sum.
\subsubsection*{(2) There is at most one $\psi_{i_s \dots i_{t+1}}$ with $s = t+1$.} \label{cancellation condition}
In this case, we can find cyclic permutation $i_1 \dots i_{r_\Gamma} \rightarrow i_{j+1} \dots i_{r_\Gamma} i_1 \dots i_j \rightarrow i_{k+1} \dots i_{r_\Gamma} \\ i_1 \dots i_k \rightarrow i_{l+1} \dots i_{r_\Gamma} i_1 \dots i_l \rightarrow i_{m+1} \dots i_{r_\Gamma} i_1 \dots i_m$ that leave the denominator of (\ref{eq:general term}) invariant. If we assume $j = k+1$ and sum over all cyclic permutations of the indices, we get an expression of the form
\begin{align}
&\frac{1}{\prod_{d\in D}\psi^2_d}
\frac{1}{\sum_{d \in D} \phi_d \prod_{\substack{d' \in D, \\  d' \neq d}} \psi_{d'}} \nonumber\\
&\times\Big[\phi_{i_{r_\gamma} \dots i_{k+2}} \psi_{i_{k+1}} \psi_{i_k \dots i_{l+1}} \psi_{i_l \dots i_{m+1}} \psi_{i_m \dots i_{1}} \nonumber\\
&\hspace{10mm}+ (\mbox{cyclic permutations})\Big]
.
\end{align}
Since the denominator of the second factor and the third factor are equal, the whole term turns to unity, and what we get is just the prefactor
\begin{align}
\frac{1}{\prod_{d\in D}\psi^2_d}
\end{align}
which is solely a function of the first Symanzik polynomial and no longer on the scattering angles.\\

In both cases the angle-dependence drops out and therefore $\Phi_R^{(1)}$ is indeed angle-independent for antisymmetric flags.

\section{A general formula} \label{sec:general formula}

The method we want to present now is in principle based on the idea to figure out all possible combinations of subgraphs building a forest and, afterwards, discard those forests that cancel in the sum. Let us look again at the graph $\gamma_{ijk}$ we already considered in section \ref{sec:L-linear term}. The forest-set of the graph is given in (\ref{eq:forest N=3}).
The terms generated by the forests are (cf. (\ref{eq:integrand N=3}))
\begin{align} \label{eq:connection forests cut-graphs}
\begin{array}{lll}
\emptyset : & 1/\psi_{ijk}^2 & \leftrightarrow \gamma_{ijk}\\
\gamma_k : & \phi_{ij} \psi_k /  \psi_{ij}^2 \psi_k^2  & \leftrightarrow \gamma_{ij} \gamma_k\\
\gamma_{jk} : & \phi_{i} \psi_{jk} / \psi_{i}^2 \psi_{jk}^2 & \leftrightarrow \gamma_i \gamma_{jk} \\
\gamma_{jk} \cup \gamma_k : & \phi_i \psi_j \psi_k / \psi_i^2 \psi_j^2 \psi_k^2 & \leftrightarrow \gamma_i \gamma_j \gamma_k
\end{array}
\end{align}
where we neglect the denominators of the second factor because they are just the sum of all different permutations of the numerator. If we forget the forests of the graph for a moment, the upper terms in the forest-sum can be regarded as they would be generated by the graphs on the right. 
In order to find a connection between the forest-set on the left and the graph-set on the right, we take the rooted tree associated with the graph and draw a box around each forest, giving us
\begin{align}\label{eq:boxes}
\vcenter{\hbox{\includegraphics[scale=0.3]{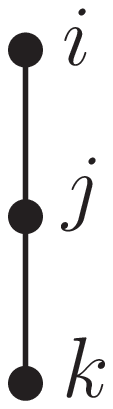}}},
\vcenter{\hbox{\includegraphics[scale=0.3]{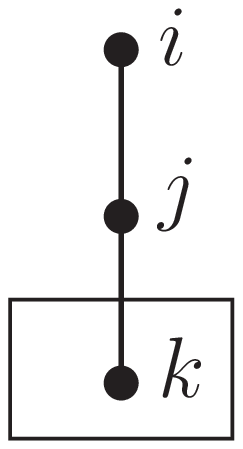}}},
\vcenter{\hbox{\includegraphics[scale=0.3]{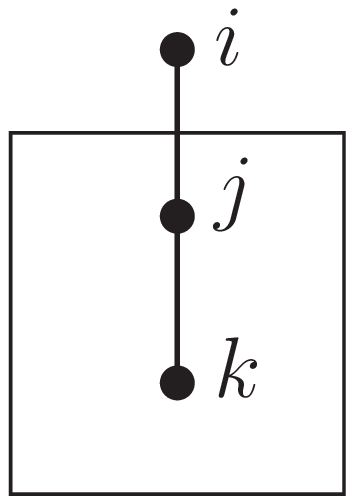}}}, \hspace{2mm}\hbox{and}\hspace{2mm}
\vcenter{\hbox{\includegraphics[scale=0.3]{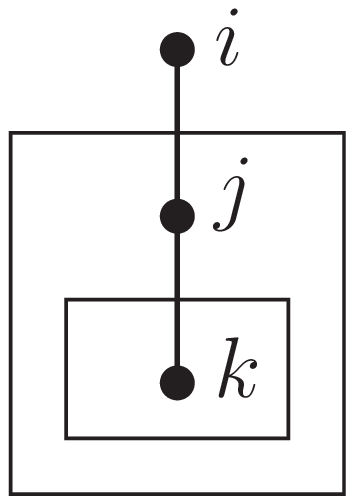}}}
\end{align} 
with the nested boxes standing for the disjoint union of the graphs within. For reasons that will become clear in a moment, we cut the trees at each edge that is crossed by a box, yielding the graphs
\begin{align}\label{eq:cut boxes}
\vcenter{\hbox{\includegraphics[scale=0.3]{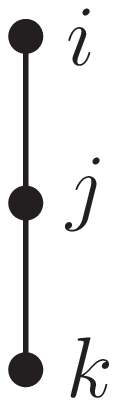}}},
\vcenter{\hbox{\includegraphics[scale=0.3]{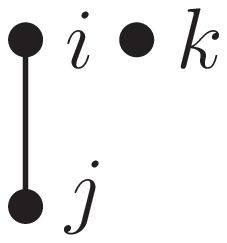}}},
\vcenter{\hbox{\includegraphics[scale=0.3]{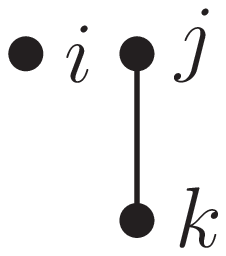}}},
\hspace{2mm} \mbox{ and } \hspace{2mm}
\vcenter{\hbox{\includegraphics[scale=0.3]{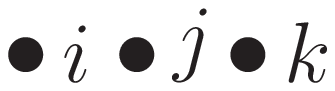}}} .
\end{align}
But cutting the tree can be understood as a contraction, splitting the graph $\gamma$ into a product of graphs $\left(\gamma/\gamma_d\right) \,\gamma_d = \gamma_r \gamma_d$ where the rooted part $\gamma_r$ is generated by contracting the original graph $\gamma$ with the dissected graph $\gamma_d$. Therefore, the upper set of dissected trees (\ref{eq:cut boxes}) in terms of graphs is given by
\begin{align} \label{eq:graphs of the cut boxes}
\gamma_{ijk}, \hspace{3mm} \gamma_{ij} \gamma_k, \hspace{3mm} \gamma_i \gamma_{jk}, 
\hspace{3mm} \mbox{ and} \hspace{3mm} \gamma_i \gamma_j \gamma_k
\end{align}
which is exactly the graph-set on the right in (\ref{eq:connection forests cut-graphs}). Moreover, (\ref{eq:cut boxes}) is the set of all partitions of the rooted tree $\lambda_3^{(ijk)} = \vcenter{\hbox{\includegraphics[scale=0.25]{graphs/box1.eps}}}$ associated with the graph $\gamma_{ijk}$. 
Thus, we can re-express the integrand (\ref{eq:linear renormalized feynman rules}) in terms of partitions of the rooted tree associated with the graph instead of forests.

To see how this can be done, we first want to reformulate the $L$-linear term of the renormalized Feynman rules (\ref{eq:linear renormalized feynman rules}) for ladder graphs.\\
Let $\mathcal{P}\left(\lambda_{r_\Gamma}^{(i_1 \dots i_{r_\Gamma})}\right)$ be the set of all partitions of the ladder $\lambda_{r_\Gamma}^{(i_1 \dots i_{r_\Gamma})}$ into $1$ up to $r_\Gamma$ ladders $\lambda_{k_i}^{(d_i)}$ with weight $k_i$ and decoration $d_i$. Of course we have $|d_i| = k_i$, otherwise this would not make sense. The set $(D, \prec)$ of all decorations $D = \left\{i_1, \dots , i_{r_\Gamma} \right\}$ has a strict total ordering $i_k \prec i_j, \hspace{2mm} \forall j, k \in \mathbb{R}: j< k$ such that the Hasse diagram of $D$ is given by the decorated rooted tree belonging to it.\\
We decompose the set $\mathcal{P}$ into subsets $\mathcal{P}^{(n)}$ fulfilling
\begin{align} \label{eq:partition set}
\mathcal{P} &= \bigcup_{n = 1}^{r_\Gamma} \mathcal{P}^{(n)} \hspace{3mm} \mbox{ and } \nonumber\\
 \mathcal{P}^{(n)} &= \left\{ p \in \mathcal{P}: p = \bigcup_{i = 1}^n \lambda_{k_i}^{(d_i)} \right.\nonumber\\ 
 &\hspace{10mm}\left. \land  \sum_{i = 1}^n k_i = r_\Gamma \land d_{i+1} \prec d_i \right\}
\end{align} 
with subsets $d_i \subseteq D$ such that $D = \bigcup_{i = 1}^n d_i$. The ordering condition $d_{i+1} \prec d_i$ means that $i_k \prec i_j, \hspace{2mm} \forall i_j \in d_i , i_k \in d_{i+1}$ and of course $i_l \prec i_j, \hspace{2mm} \forall i_j, i_l \in d_i: j < l$. Thus, the order of the decorations always has to stay the same no matter how many dissections were performed.
For example, the set (\ref{eq:cut boxes}) can be written as 
\begin{gather}
\mathcal{P} \left(\lambda_3^{(ijk)}\right) = \mathcal{P}^{(1)} \cup \mathcal{P}^{(2)} \cup \mathcal{P}^{(3)}\hspace{5mm} \mbox{ with } \nonumber\\
\mathcal{P}^{(1)} = \left\{\lambda_3^{(ijk)} \right\}, \hspace{3mm}
\mathcal{P}^{(2)} = \left\{\lambda_2^{(ij)} \lambda_1^{(k)}, \lambda_1^{(i)} \lambda_2^{(jk)} \right\}, \nonumber\\
 \mbox{ and } \hspace{3mm} \mathcal{P}^{(3)} = \left\{\lambda_1^{(i)} \lambda_1^{(j)} \lambda_1^{(k)}\right\} .
\end{gather}
%
%
Since each element of $\mathcal{P} \left(\lambda_{r_\Gamma}^{(D)}\right)$ can be linked to a  
term in $\Phi_R^{(1)}\left(\gamma_D\right)$ of the graph $\gamma_D$, 
whose subgraph structure is visualized through $\lambda_{r_\Gamma}^{(D)}$, we can reformulate $\Phi_R^{(1)} \left(\lambda_{r_\Gamma}^{(D)}\right)$ in terms of ladders as follows
\begin{align} \label{eq:renormalized linear term for ladders}
&\Phi_R^{(1)} \left(\gamma_{i_1 \dots i_{r_\Gamma}}\right) \equiv 
	\Phi_R^{(1)} \left(\lambda_{r_\Gamma}^{(i_1 \dots i_{r_\Gamma})}\right) = \nonumber\\
&\hspace{3mm} \sum_{n = 1}^{r_\Gamma} \left(-1\right)^{n + 1} \hspace{-5mm}\sum_{\substack{p \in \mathcal{P}^{(n)}(\lambda_{r_\Gamma}^{(D)}) \\ p = \cup_{i=1}^n \lambda_{k_i}^{(d_i)}}} \Bigg[\frac{1}{\prod_{i=1}^n \psi_{d_i}^2} \nonumber\\
&\hspace{28mm}\times \frac{\phi_{d_1} \prod_{i=2}^n \psi_{d_i}}{\sum_{i = 1}^n \phi_{d_i}\prod_{\substack{j = 1, \\ j \neq i}}^n \psi_{d_j}}\Bigg].
\end{align}
For convenience, we make use of the notation $\psi_{d_i} = \psi \left(\lambda_{k_i}^{(d_i)}\right) = \psi \left(\gamma_{d_i}\right)$ and analogously $\phi_{d_i}$.
\\
Now we go one step further and ask for $\Phi_R^{(1)}\left(\Lambda_{r_\Gamma}^-\right)$ in terms of the possible partitions $p\in \mathcal{P}$. Therefore, we have to recall the cases (1) and (2) from section \ref{sec:L-linear term} for the integrand of $\Phi_R^{(1)}\left(\Lambda_{r_\Gamma}^-\right)$. From case (1) it follows in terms of partitions that all elements $p \in \mathcal{P}$ containing more than one ladder of weight $1$ do not contribute to the integrand, as we have seen for the fourth forest in (\ref{eq:cut boxes}). The second case tells us that the sum of all partitions of a graph consisting of the same number of ladders with fixed weight, contributes only one single term to the integrand (cf. the second and third forest in (\ref{eq:cut boxes})).
\\
To get a more adapted formulation of those two cases, we define a multiplicity $m_p(k)$ for each element $p$ in (\ref{eq:partition set}), given by the number of ladders $\lambda_{k_i = k}^{(d_i)}$ of weight $k$ contained in $p$. Let $m_p$ be the $r_\Gamma$-tuple of all multiplicities of $p$, i.e. $m_p = \left(m_p(k)\right)_{k=1,\dots ,r_\Gamma} = \left(m_p(1), \dots, m_p(r_\Gamma)\right)$. Then, we claim that two elements $p$ and $p'$ of $\mathcal{P}$ are independent of each other if and only if there exists at least one weight $k$ such that $m_p(k) \neq m_{p'}(k)$. That is to say that $p$ and $p'$ have not the same tuple of multiplicities. Therefore, it follows that the two forests $\vcenter{\hbox{\includegraphics[scale=0.3]{graphs/cut-boxes1.eps}}}$ and $\vcenter{\hbox{\includegraphics[scale=0.3]{graphs/cut-boxes2.eps}}}$ are not independent of each other since they have the same multiplicity tuple given by $\left(m(1) = 1, m(2) = 1, m(3) = 0\right)$.\\
Based on the invented notion of the multiplicity tuple, we define
\begin{align}\label{eq:independent partition set multiplicity}
\mathcal{P}_{\tiny\mbox{ind}}\left(\lambda_{r_\Gamma}^{(D)}\right) \coloneqq &
\left\{p\in \mathcal{P}\left(\lambda_{r_\Gamma}^{(D)}\right):\right. \nonumber\\
&\hspace{-15mm}\left. m_p \neq m_{p'} \, \forall \, p, p' \in \mathcal{P}_{\tiny\mbox{ind}}\left(\lambda_{r_\Gamma}^{(D)}\right)  \right\}
\end{align}
 to be the set of all independent partitions of $\lambda_{r_\Gamma}^{(D)}$, which means that all elements of $\mathcal{P}_{\tiny\mbox{ind}}$ are pairwise independent of each other. Clearly, this set is not unique because out of all partitions in $\mathcal{P}\left(\lambda_{r_\Gamma}^{(D)}\right)$ with the same multiplicity tuple we have to choose only one to be contained in $\mathcal{P}_{\tiny\mbox{ind}}$. Nevertheless, this will not bother us since our formula is completely independent of the choice of the specific partition. \\
Consequently, without loss of generality, we will always choose the partition consisting of ladders with equal or decreasing weight for a fixed multiplicity tuple, meaning that $p = \cup_i \lambda_{k_i}^{(d_i)}$ with $k_i \geq k_{i+1} \geq 1$. This selection rule allows us to write the set of independent partitions of a ladder $\lambda_{r_\Gamma}^{(D)}$ as
\begin{align} 
\mathcal{P}_{\tiny \mbox{ind}} &= \bigcup_{n = 1}^{r_\Gamma} \mathcal{P}_{\tiny \mbox{ind}}^{(n)} \hspace{3mm}  \mbox{with} \nonumber\\
 \mathcal{P}_{\tiny \mbox{ind}}^{(n)} &= \Bigg\{ p \in \mathcal{P}_{\tiny \mbox{ind}}: p = \bigcup_{i = 1}^n \lambda_{k_i}^{(d_i)} \land  \sum_{i = 1}^n k_i = r_\Gamma  \nonumber\\
 &\hspace{10mm} \land k_i \geq k_{i+1} \geq 1\land d_{i+1} \prec d_i \Bigg\} .
\label{eq:independent partition set}
\end{align} 
Now we are on the verge of giving a compact formulation of the final integrand $\Phi_R^{(1)}\left(\Lambda_{r_\Gamma}^-\right)$. Calling to mind that all partitions containing more than one ladder of weight $1$ do not contribute to the integrand (cf. case (1) in section \ref{sec:L-linear term}),  $\mathcal{P}_{\tiny\mbox{ind}}\left(\lambda_{r_\Gamma}^{(D)}\right)$ gives the full set of partitions we need to build up the integrand of $\Lambda_{r_\Gamma}^-$ if we discard all elements $p$ with $m_p(1) > 1$. The only thing we need for the final expression is the numerical prefactor of the terms contributing to the integrand. Let us see how this can be done by giving an example.
Consider the case $r_\Gamma = 6$. The set of partitions $p$ with multiplicity tuple $m_p = (1,1,1)$ is given by
\begin{gather}
\vcenter{\hbox{\includegraphics[scale=0.23]{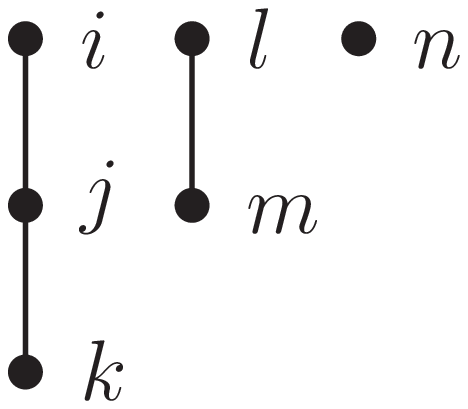}}}, \hspace{2mm}
\vcenter{\hbox{\includegraphics[scale=0.23]{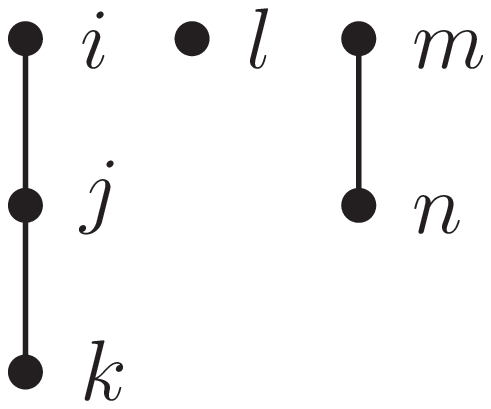}}}, \hspace{2mm}
\vcenter{\hbox{\includegraphics[scale=0.23]{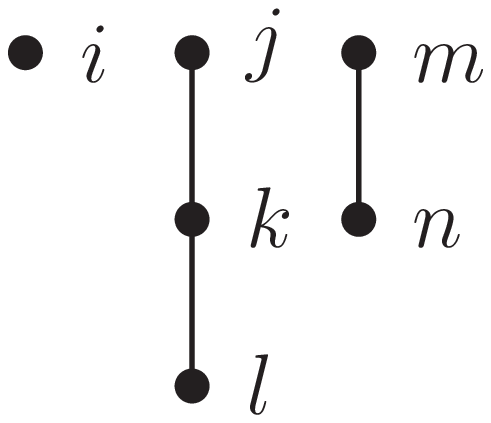}}}, \hspace{2mm}
\vcenter{\hbox{\includegraphics[scale=0.23]{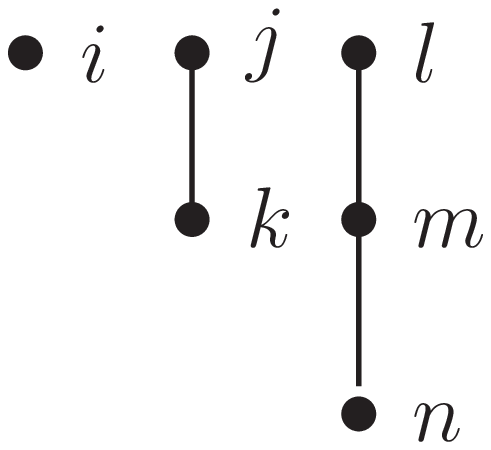}}}, \nonumber\\
\vcenter{\hbox{\includegraphics[scale=0.23]{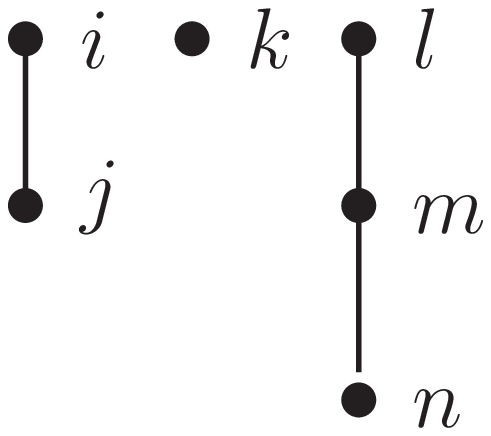}}}, \hspace{1mm}\mbox{ and}\hspace{2mm}
\vcenter{\hbox{\includegraphics[scale=0.23]{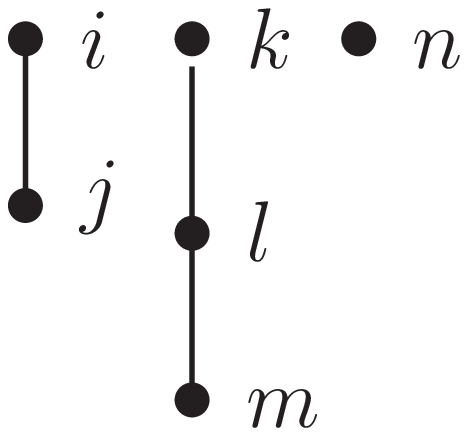}}} .
\end{gather}
As we already know, those terms will sum up to the single one $\frac{1}{\psi_{ijk}^2 \psi_{lm}^2\psi_n^2}$ times a numerical prefactor which is composed as follows. The number of partitions with the same multiplicity for $r_\Gamma$ fixed is just the number of possibilities to arrange the ladders in the respective partition. If we forget the decorations of the trees for a moment, this number would be just the number of ladders in $p$ factorial. Since the decorations of the ladders are strictly ordered we do not get a new partition if we interchange ladders of the same weight in the partition. Therefore, the factorial of the number of ladders in $p$ has to be divided by the multiplicity factorial for each weight. In our example we get $\frac{3!}{1! 1! 1!} = 6$ for the number of partitions with multiplicity $m_p = (1,1,1)$. The integrand associated to the first partition in the upper example is (cf. equation (\ref{eq:renormalized linear term for ladders})) $\frac{1}{\psi_{ijk}^2 \psi_{lm}^2\psi_n^2} \frac{\phi_{ijk} \psi_{lm} \psi_n}{\phi_{ijk} \psi_{lm} \psi_n + \phi_{lm} \psi_{ijk} \psi_n + \phi_n \psi_{ijk} \psi_{lm}}$. As one can see, each three of the partitions add up in such a way that the angle dependent term turns to unity. More generally, if $n$ is the number of ladders in the partition, each $n$ of the partitions will add up to one term in the sum. Therefore, dividing the number of partitions with the same multiplicity by the number of ladders in each partition gives the prefactor in the integrand we are looking for. In the present example this factor is $\frac{3!}{1!1!1! 3} = 2$. In general, the prefactor can be defined as
\begin{align} \label{eq:prefactor}
\frac{\frac{n_p !}{\prod_{i=1}^{r_\Gamma} m_p(k_i)}}{n_p} &= \frac{(n_p - 1) !}{\prod_{i=1}^{r_\Gamma} m_p(k_i)!} \nonumber\\
&= \frac{\left(\sum_{j=1}^{r_\Gamma} m_p(k_j) - 1\right)!}{\prod_{i=1}^{r_\Gamma} m_p(k_i)!}
\end{align}
with $n_p$ the number of ladders in the partition $p$. For an example of the prefactor for some partitions see table \ref{tab:prefactor}, in which we omitted the factors $0!$ in the denominators and the decorations of the trees for convenience. 
\begin{table}[H]
\centering
\begin{tabular}{|>{$}l<{$}|>{$}r<{$}>{$}c<{$}>{$}l<{$}|}
\hline
\mbox{Partition} \hspace{3mm} p & \multicolumn{3}{>{$}c<{$}|}{\mbox{Prefactor} \hspace{3mm}  \frac{(n_p-1)!}{\prod_i m_p(k_i)!}} \\
\hline
\vcenter{\hbox{\includegraphics[scale=0.2]{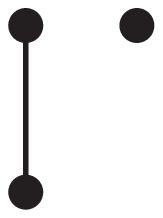}}} & \frac{1!}{1! 1!} &=& 1 \\
\vcenter{\hbox{\includegraphics[scale=0.2]{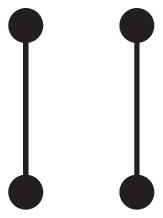}}} & \frac{1!}{2!} &=& \frac{1}{2} \\
\vcenter{\hbox{\includegraphics[scale=0.2]{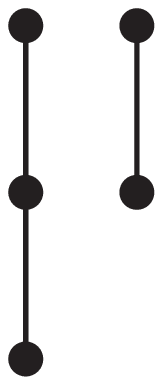}}} & \frac{1!}{1! 1!} &=& 1 \\
\vcenter{\hbox{\includegraphics[scale=0.2]{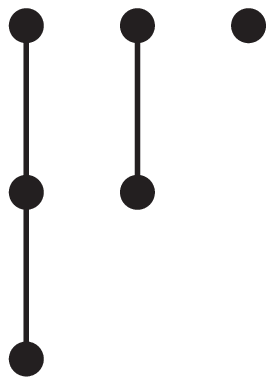}}} & \frac{2!}{1!1!1!} &=& 2 \\
\vcenter{\hbox{\includegraphics[scale=0.2]{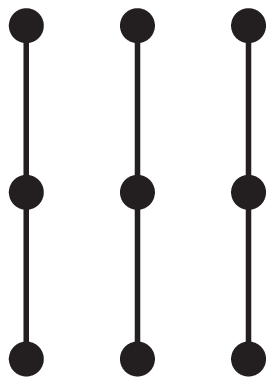}}} & \frac{2!}{3!} &=& \frac{1}{3} \\
\vcenter{\hbox{\includegraphics[scale=0.2]{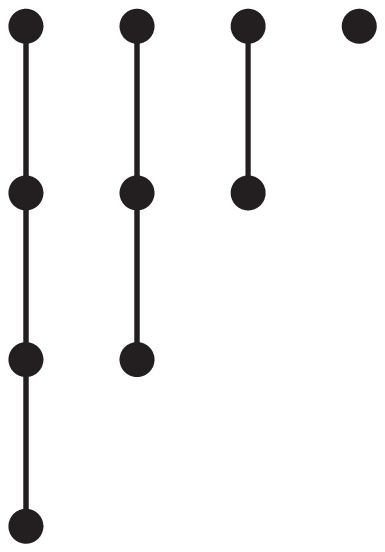}}} & \frac{3!}{1!1!1!1!} &=& 6 \\
\vcenter{\hbox{\includegraphics[scale=0.2]{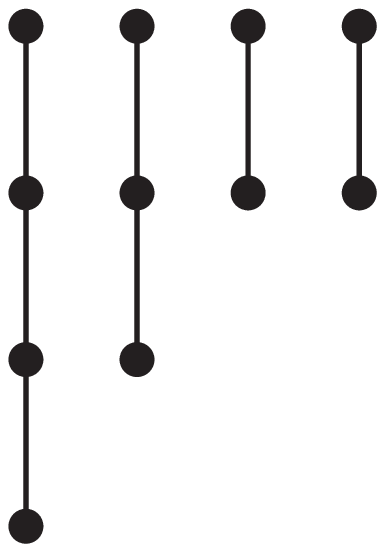}}} & \frac{3!}{2!1!1!} &=& 3 \\
\vcenter{\hbox{\includegraphics[scale=0.2]{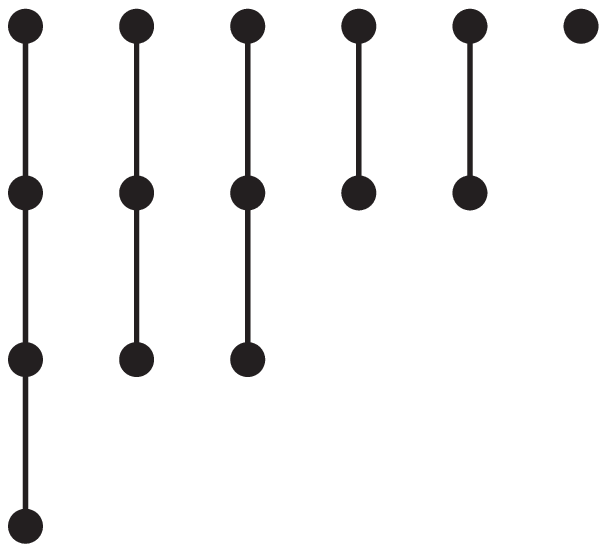}}} & \frac{5!}{1!2!2!1!} &=& 30 \\
\hline
\end{tabular}
\caption{The prefactor from equation (\ref{eq:prefactor}) for a variety of partitions.}
\label{tab:prefactor}
\end{table}
Finally, the linear term of the renormalized Feynman rules for an antisymmetric flag takes the compact form
\begin{align} \label{eq:linear term antisymmetric flag}
&\Phi_R^{(1)} \left(\Lambda_{r_\Gamma}^-\right) = \nonumber\\
&\hspace{3mm}\int_\gamma \Omega_\gamma \sum_{n = 1}^{r_\Gamma} \left(-1\right)^{n+1} 
\sum_{d \in D} \sum_{d = 1}^{r_\Gamma} \tilde{\varepsilon}_{(D)} \nonumber\\
&\hspace{10mm}\sum_{\substack{p \in \mathcal{P}_{\tiny\mbox{ind}}^{(n)}(\lambda_{r_\Gamma}^{(D)}) \\
								p = \cup_{i=1}^n \lambda_{k_i}^{(d_i)} \\
								m_p(1) \leq 1}}
\frac{(n-1)!}{\prod_{j=1}^{r_\Gamma} m_p(k_j)!} \frac{1}{\prod_{i=1}^n \psi_{d_i}^2}
\end{align}
%
where $\tilde{\varepsilon}_{(D)}$ is the tensor defined in (\ref{eq:epsilon tilde}) indexed by the full decoration set of the corresponding flag and
the sum $\sum_{d \in D} \sum_{d = 1}^{r_\Gamma}$ runs over all elements of the decoration set, each of them taking on values from $1$ up to $r_\Gamma$. Note, that $d$ denotes an element and $d_i$ an ordered subset of $D$ so they should not be confused with each other.\\
The upper result is very striking since the problem of finding all the forests of a graph is boiled down to the task of finding all partitions of the corresponding ladder graph which is straightforward. A very easy and pictorial way to cope with this task will be given in the following section.

\section{A pictorial approach using flag diagrams} \label{sec:pictorial approach}

\begin{figure}[H]
\begin{center}
\subfigure[Ferrers diagrams for $r_\Gamma = 3$]{
\includegraphics[scale=0.4]{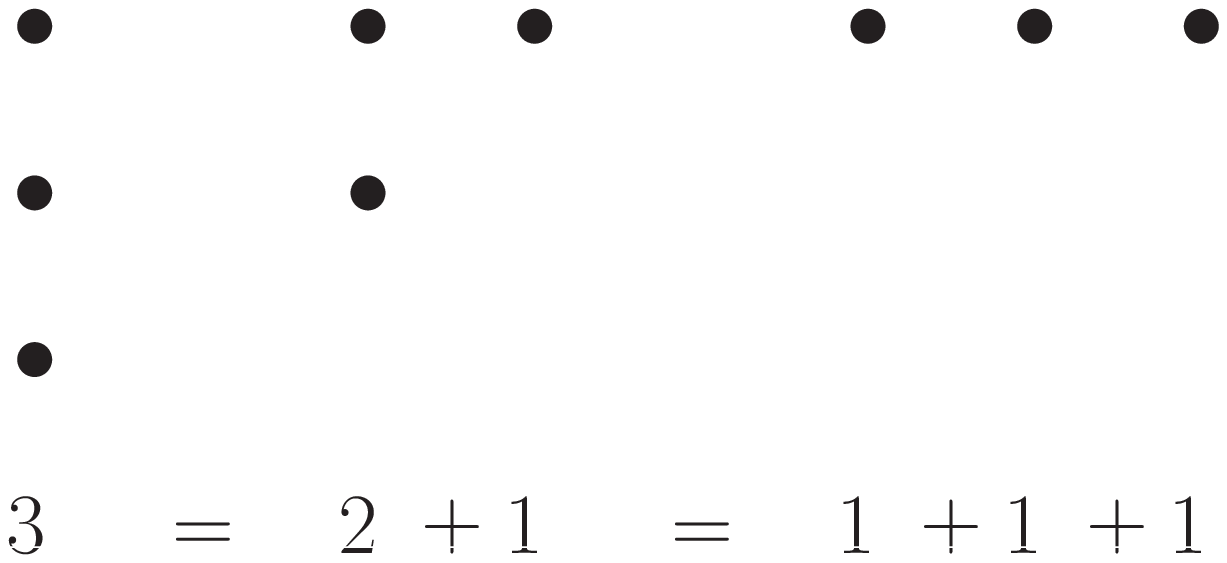}
\label{subfig:ferrers diagram r=3}
} 
${\footnotesize\mbox{label vertices}} \Bigg\Downarrow {\footnotesize\mbox{draw edges}}$
\subfigure[Flag diagram for $r_\Gamma = 3$]{
\includegraphics[scale=0.4]{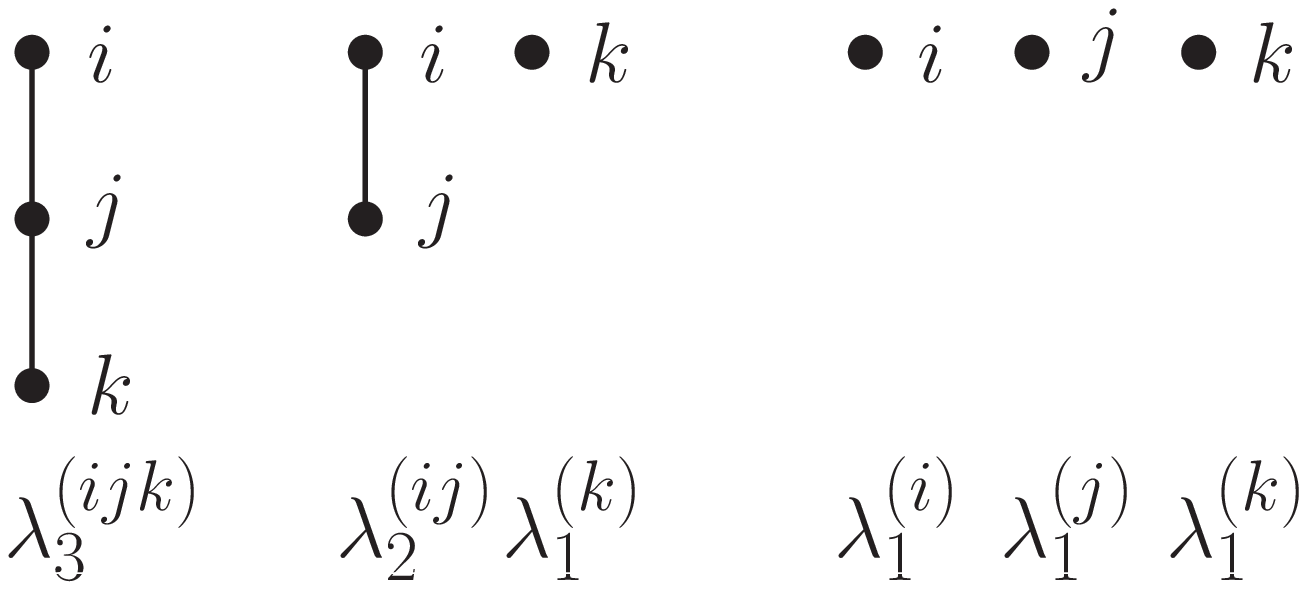}
\label{subfig:flag diagram r=3}
}
\caption{Ferrers diagrams of all possible partitions of $r_\Gamma = 3$ into positive integers and the corresponding flag diagram constructed out of them.}
\label{fig:ferrers + flag diagram r=3}
\end{center}
\end{figure}

In the last section we presented a formula that allows us to calculate the linear term of the renormalized Feynman rules of an antisymmetric flag by looking at the possible (independent) partitions of the corresponding ladder.
There is also a pictorial way to deduce the set $\mathcal{P}_{\tiny\mbox{ind}}\left(\lambda_{r_\Gamma}^{(D)}\right)$ for a given ladder $\lambda_{r_\Gamma}^{(D)}$, which is based on the idea of Ferrers diagrams (see appendix \ref{app:ferrers diagram}). Such a pictorial representation of the set of independent partitions of a ladder graph, we will refer to as a flag diagram. To see how these diagrams can be constructed, we first 
look at the set of all independent partitions of a ladder (see equation (\ref{eq:independent partition set})). If we ignore the decorations of the ladders for a moment, the partition-set is defined by the condition
\begin{align}\label{eq:partition coradical degree}
 r_\Gamma = \sum_{i=1}^n k_i  \hspace{2mm} \mbox{ and } \hspace{2mm} k_i \geq k_{i+1} \geq 1 .
\end{align}
Obviously, equation (\ref{eq:partition coradical degree}) defines the set of all partitions of the co-radical degree $r_\Gamma$ into a sum of $n$ positive integers $k_i$. This partition can be illustrated by drawing the corresponding Ferrers diagram.\\
Consider the case $r_\Gamma = 3$. The possible decompositions of $r_\Gamma$ are $r_\Gamma = 3 = 2 + 1 = 1 + 1 + 1$ with the corresponding Ferrers diagrams given in figure \ref{subfig:ferrers diagram r=3}. Now, the set of independent partitions of the corresponding ladder $\lambda_{r_\Gamma = 3}^{(ijk)}$ can directly be constructed out of Ferrers diagrams in figure \ref{subfig:ferrers diagram r=3} by simply drawing edges between the dots in the same column. Afterwards, we can label the dots in each diagram by the elements in the decoration set, going from top to bottom and from left to right, yielding the flag diagram in figure \ref{subfig:flag diagram r=3}. The generalization of this construction is straightforward.
To determine the partitions in the flag diagram that contribute to the integrand of $\Phi_R^{(1)}\left(\Lambda_{r_\Gamma}^-\right)$, we have to discard all partitions with more than one ladder of weight one, e.g. the third one in figure \ref{subfig:flag diagram r=3}. Or, in terms of Ferrers diagrams, we discard all diagrams containing more than one column with only one entry. Figure \ref{fig:flag diagrams r=4,5} shows the flag diagram for co-radical degree four and five, in which we already crossed out those partitions that do not show up in the integrand.
\begin{figure}[H]
\begin{center}
\subfigure[Flag diagram for $r_\Gamma = 4$]{
\includegraphics[scale=0.25]{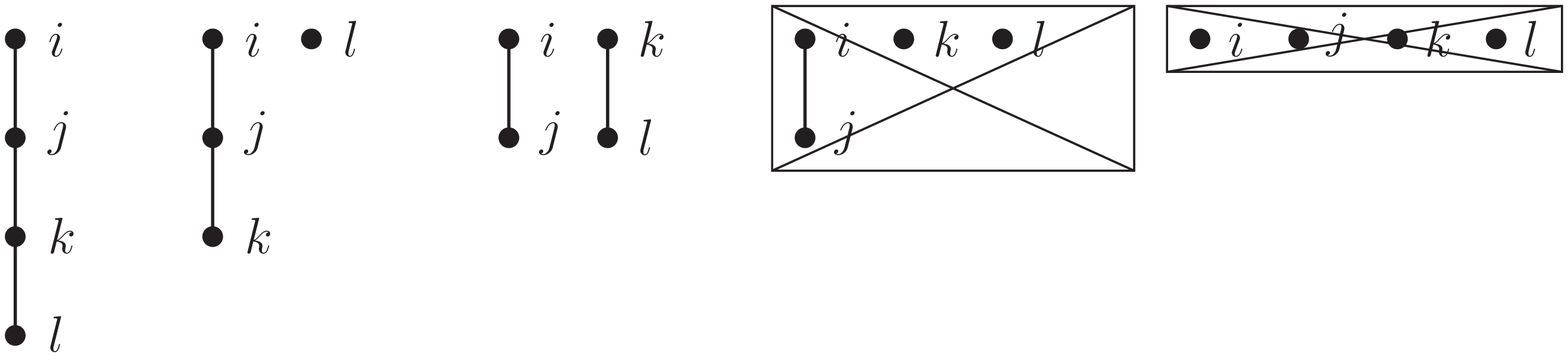}
\label{subfig:flag diagram r=4}
} 
\subfigure[Flag diagram for $r_\Gamma = 5$]{
\includegraphics[scale=0.25]{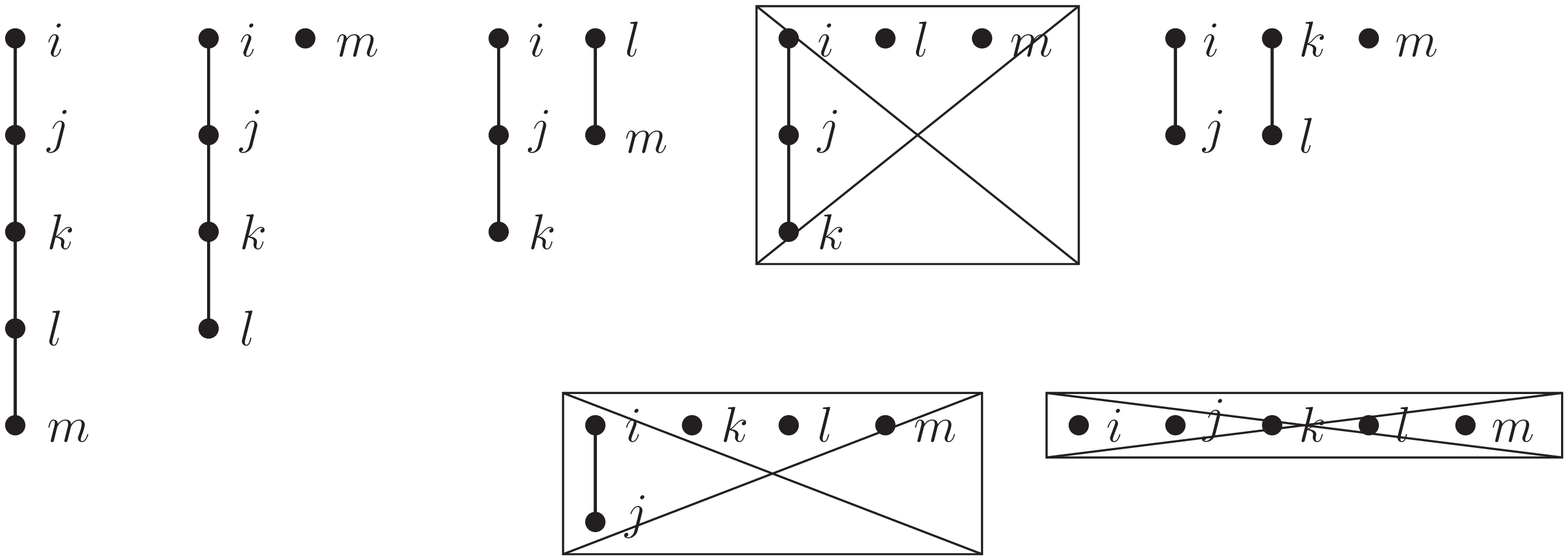}
\label{subfig:flag diagram r=5}
}
\caption{Flag diagrams for co-radical degree four and five with all partitions that do not contribute to integrand crossed out.}
\label{fig:flag diagrams r=4,5}
\end{center}
\end{figure}

\section{Example: $r_\Gamma = 6$} \label{sec:example}

\begin{figure}[H]
\centering
\includegraphics[scale=0.2]{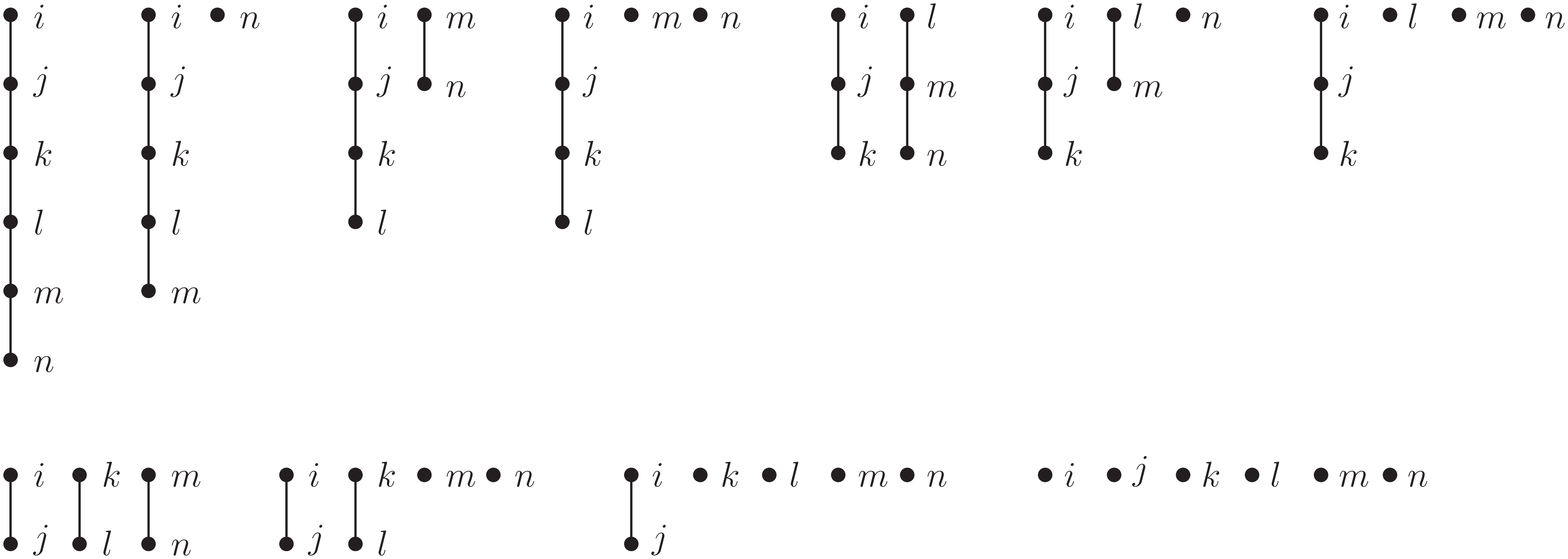}
\caption{Flag diagram for co-radical degree $r_\Gamma = 6$.}
\label{fig:flag diagram r=6}
\end{figure}

We consider the case $r_\Gamma = 6$. There are $11$ different possibilities to decompose $6$ into a sum of positive integers, namely 
\begin{align}
6 &= 5 + 1 = 4 + 2 = 4 + 1 + 1 = 3 + 3 \nonumber\\
&= 3 + 2 + 1 = 3 + 1 + 1 + 1 = 2 + 2 + 2  \nonumber\\
&= 2 + 2 + 1 + 1 = 2 + 1 + 1 + 1 + 1 \nonumber\\
&= 1 + 1 + 1 + 1 + 1 + 1 .
\end{align}
Each of those decompositions can be illustrated by a Ferrers diagram, and from the set of diagrams we can deduce the corresponding flag diagram given in figure \ref{fig:flag diagram r=6}. 
Our goal is to calculate the integrand of the antisymmetric flag. Therefore, we can discard the fourth and seventh partition in the first, and all but the first partition in the second line of the flag diagram. The remaining partitions that contribute to the integrand are
\begin{gather}
\lambda_6^{(ijklmn)} , \hspace{1mm}
\lambda_5^{(ijklm)} \lambda_1^{(n)} , \hspace{1mm}
\lambda_4^{(ijkl)} \lambda_2^{(mn)} , \hspace{1mm}
\lambda_3^{(ijk)} \lambda_3^{(lmn)} , \nonumber\\ 
\lambda_3^{(ijk)} \lambda_2^{(lm)} \lambda_1^{(n)}, \hspace{1mm}
\mbox{and} \hspace{1mm}
\lambda_2^{(ij)} \lambda_2^{(kl)} \lambda_2^{(mn)}
\end{gather}
with prefactors (see equation (\ref{eq:prefactor}))
\begin{align}
1 , \hspace{3mm}
1 , \hspace{3mm}
1 , \hspace{3mm}
\frac{1}{2} , \hspace{3mm}
2 , \hspace{5mm} \mbox{and} \hspace{3mm}
\frac{1}{3} .
\end{align}
Thus, the integrand from equation (\ref{eq:linear term antisymmetric flag}) evaluates to
\begin{align} \label{eq:linear term r=6}
\Phi_R^{(1)}& \left(\Lambda_6^-\right) = \int \Omega \hspace{-2mm}\sum_{i,j,k,l,m,n = 1}^6 \hspace{-1mm} \tilde{\varepsilon}_{ijklmn} \Bigg[
\frac{1}{\psi_{ijklmn}^2}  \nonumber\\
&\hspace{-5mm}
- \frac{1}{\psi_{ijklm}^2 \psi_n^2} 
- \frac{1}{\psi_{ijkl}^2 \psi_{mn}^2} 
- \frac{1}{2} \frac{1}{\psi_{ijk}^2\psi_{lmn}^2}\nonumber\\
&\hspace{11mm}
+ \frac{2}{\psi_{ijk}^2 \psi_{lm}^2 \psi_n^2}
+ \frac{1}{3} \frac{1}{\psi_{ij}^2 \psi_{kl}^2 \psi_{mn}^2}
\Bigg] .
\end{align}
This calculation was done without any great effort and nearly took a half page. In contrast, we can think of the explicit calculation: For the ladder of co-radical degree $6$ we would get a total number of $32$ forests, each giving us one term in the integrand corresponding to the forest formula in equation (\ref{eq:linear renormalized feynman rules}). To see how the angle-dependence cancels out, we can rewrite the integrand using $\tilde{\varepsilon}_{ijklmn} = \frac{1}{6} \left[\tilde{\varepsilon}_{ijklmn} + \tilde{\varepsilon}_{jklmni} + \dots\right]$, giving us a total amount of $32 \times 6 = 192$ terms that have to be combined until we end up with the expression (\ref{eq:linear term r=6}). Thus, our formula not only saves a lot of time and paperwork but also is more elegant in a combinatorial sense.

\section{Conclusion}

Within the scope of the present article, we initially considered scalar Feynman integrals in parametric representation. After introducing dimensionless scattering angles and carrying out one of the integrations, it becomes apparent that the renormalized Feynman rules can be written as a polynomial in the scaling parameter $L = \ln\left(S/S_0\right)$. Assuming that $L$ is very small, the dominant contributions of $\Phi_R$ arise from the low-order terms in the polynomial. Therefore, we focus on the $L$-linear term of the renormalized Feynman rules.
In quest of finding combinations of graphs such that the linear term is significantly simplified, we end up at so-called flags.
It turned out that in the case of antisymmetric flags (as well as in the case of symmetric flags) the $\phi$-dependence and thus the angle-dependence drops out in the linear term of $\Phi_R$ if we assume that scattering angles are preserved.
Based on our discovery, we thought about a concept how to compute $\Phi_R^{(1)}$ for antisymmetric flags of arbitrary order and developed a formula whereby the calculations are facilitated and shortened. In a concluding example, we showed how advantageous this formula can be.
Indeed, the formula can also be used to compute $\Phi_R^{(1)}$ for symmetric flags. The sole difference is that all ladders in the flag-diagram contribute to the integrand. Consequently, the claim $m_p(1) \leq 1$ is dropped (cf. equation (\ref{eq:linear term antisymmetric flag})). Apart from this, the formula is unchanged such as the computation of the prefactor (cf. equation (\ref{eq:prefactor})).
Within the context of this article, we also considered combinations of ladder graphs with branched rooted trees for $r_\Gamma$ fixed. However, investigations of those combinations of graphs gave no reason to expect the angle-independence of the $L$-linear term so far.

\appendix
\setcounter{equation}{0}
\renewcommand{\theequation}{\Alph{section}.\arabic{equation}}  
\setcounter{defi}{0}
\renewcommand{\thedefi}{\Alph{section}.\arabic{defi}}
\section{Algebras} \label{app:algebras}
In the following we want to give some basic definitions regarding algebras.
What we aim at with this appendix is to give a brief overview of this topic and not a full mathematical description.\\
Let $\mathbb{K}$ be a field of characteristic zero, $V_1$ and $V_2$ two vector spaces, and $\tau_{V_1, V_2}: V_1 \otimes V_2 \rightarrow V_2 \otimes V_1$ the flip map that interchanges the elements in a tensor product $\tau(v_1\otimes v_2) = v_2 \otimes v_1$. 
\begin{defi}[Algebra]\label{def:algebra}
An associative $\mathbb{K}$-algebra ($A, m$) is a $\mathbb{K}$-vector space $A$ together with a linear map $m: A \otimes A \rightarrow A$, called product, such that
\begin{align} \label{eq:algebra condition 1}
m \circ (\operatorname{id} \otimes m) = m \circ (m \otimes \operatorname{id}).
\end{align}
If there exists a linear map $\mathbb{I}: \mathbb{K} \rightarrow A$ fulfilling 
\begin{align} \label{eq:algebra condition 2}
m \circ (\operatorname{id} \otimes \mathbb{I}) = \operatorname{id} =  m \circ (\mathbb{I} \otimes \operatorname{id}),
\end{align}
the algebra ($A, m, \mathbb{I}$) is said to be unital, and $\mathbb{I}$ is called the unit map. \\ For $m \circ \tau = m$ the algebra is commutative.
\end{defi} 
The conditions (\ref{eq:algebra condition 1}) and (\ref{eq:algebra condition 2}) are the same as demanding that the diagrams 
\begin{eqnarray} \label{fig:algebra diagram}
\vcenter{\hbox{
\begin{tikzpicture}
\matrix(M) [matrix of math nodes, row sep=2em, column sep=3em, text height=1ex, text depth=0.25ex]
{A \otimes A \otimes A & A \otimes A \\
A \otimes A & A \\} ;
\path[->]
(M-1-1) edge node[auto]{\small $m \otimes \operatorname{id}$} (M-1-2)
		edge node[auto,swap]{\small $\operatorname{id}\otimes m$} (M-2-1)
(M-2-1) edge node[auto]{\small $m$} (M-2-2)
(M-1-2) edge node[auto]{\small $m$} (M-2-2);
\end{tikzpicture}
}}
\end{eqnarray}
and
\begin{eqnarray}
\vcenter{\hbox{
\begin{tikzpicture}
\matrix(M) [matrix of math nodes, row sep=2em, column sep=3em, text height=1ex, text depth=0.25ex]
{\mathbb{K} \otimes A & A \otimes A & A \otimes \mathbb{K} \\
 & A & \\} ;
\path[->]
(M-1-1) edge node[auto]{\small $\mathbb{I} \otimes \operatorname{id}$} (M-1-2)
		edge node[auto,swap]{\small $\cong$} (M-2-2)
(M-1-3) edge node[auto,swap]{\small $\operatorname{id}\otimes\mathbb{I}$} (M-1-2)
		edge node[auto]{\small $\cong$} (M-2-2)
(M-1-2) edge node[auto]{\small $m$} (M-2-2);
\end{tikzpicture}
}}
\end{eqnarray}
commute. By reversing the arrows of the diagrams, one can derive objects which are somehow dual to algebras, namely coalgebras.
\begin{defi}[Coalgebra] \label{def:coalgebra}
A coassociative $\mathbb{K}$-coalgebra ($C, \Delta$) consists of a $\mathbb{K}$-vector space $C$ and a linear map $\Delta: C\rightarrow C\otimes C$, called coproduct, such that coassociativity is fulfilled
\begin{align} \label{eq:coalgebra condition 1}
(\operatorname{id} \otimes \Delta) \circ \Delta = (\Delta \otimes \operatorname{id}) \circ \Delta .
\end{align}
If there exists a linear map $\hat{\mathbb{I}}: C \rightarrow \mathbb{K}$ with
\begin{align} \label{eq:coalgebra condition 2}
\left(\hat{\mathbb{I}} \otimes \operatorname{id}\right) \circ \Delta = \operatorname{id} = \left(\operatorname{id} \otimes \hat{\mathbb{I}}\right) \circ \Delta ,
\end{align}
the coalgebra ($C, \Delta, \hat{\mathbb{I}}$) is said to be counital, and $\hat{\mathbb{I}}$ is called the counit map.\\
For $\tau \circ \Delta = \Delta$ the coalgebra is cocommutative.
\end{defi}
As we mentioned before, the properties (\ref{eq:coalgebra condition 1}) and (\ref{eq:coalgebra condition 2}) are equivalent to the commutativity of the diagrams
\begin{eqnarray} \label{fig:coalgebra diagram}
\vcenter{\hbox{
\begin{tikzpicture}
\matrix(M) [matrix of math nodes, row sep=2em, column sep=3em, text height=1ex, text depth=0.25ex]
{C & C \otimes C \\
C \otimes C & C \otimes C \otimes C \\} ;
\path[->]
(M-1-1) edge node[auto]{\small $\Delta$} (M-1-2)
		edge node[auto,swap]{\small $\Delta$} (M-2-1)
(M-2-1) edge node[auto]{\small $\Delta \otimes \operatorname{id}$} (M-2-2)
(M-1-2) edge node[auto]{\small $\operatorname{id} \otimes \Delta$} (M-2-2);
\end{tikzpicture}
}}
\end{eqnarray}
and
\begin{eqnarray}
\vcenter{\hbox{
\begin{tikzpicture}
\matrix(M) [matrix of math nodes, row sep=2em, column sep=3em, text height=1ex, text depth=0.25ex]
{\mathbb{K} \otimes C & C \otimes C & C \otimes \mathbb{K} \\
 & C & \\} ;
\path[->]
(M-2-2) edge node[auto]{\small $\cong$} (M-1-1)
		edge node[auto,swap]{\small $\cong$} (M-1-3)
		edge node[auto,swap]{\small $\Delta$} (M-1-2)
(M-1-2) edge node[auto,swap]{\small $\hat{\mathbb{I}} \otimes \operatorname{id}$} (M-1-1)
		edge node[auto]{\small $\operatorname{id}\otimes \hat{\mathbb{I}}$} (M-1-3);
\end{tikzpicture}
}}
\end{eqnarray}
which are dual to those in (\ref{fig:algebra diagram}). More generally, we will extend the definition of the coproduct to that of the iterated coproduct $\Delta^n: C \otimes C^{\otimes (n+1)}$ by
\begin{gather}
\Delta^0 \coloneqq \operatorname{id} \hspace{1mm}\mbox{ and } \nonumber\\
\Delta^{n+1} \coloneqq \left( \Delta \otimes \operatorname{id}^{\otimes n}\right) \circ \Delta^n \mbox{ for } n \in \mathbb{N}_0 .
 \label{eq:iterated coproduct1}
 \end{gather}
Clearly, the recursive definition above is invariant under a variation of the order in which the coproduct is applied. This fact follows from the coassociativity of $\Delta$ (see equation (\ref{eq:coalgebra condition 1})) and therefore
\begin{align}\label{eq:iterated coproduct2}
\Delta^{n+1} = \left( \operatorname{id}^{\otimes m} \otimes \Delta \otimes \operatorname{id}^{\otimes (n-m)}\right) \circ \Delta^n \nonumber\\
 \forall m, n \in \mathbb{N}_0, \, m\leq n.
\end{align}
\begin{rem}
In literature it is very common to use Sweedlers notation for the coproduct $\Delta (x) = x' \otimes x''$ with $x \in C$, which is shorthand for $\Delta (x) = \sum_i x'_{(i)} \otimes x''_{(i)}$.
\end{rem}
\begin{defi}[Algebra and coalgebra morphism]
Consider two algebras ($A_1, m_1$) and ($A_2, m_2$). The linear map $\phi: A_1 \rightarrow A_2$ is an algebra morphism if 
\begin{align}
\phi \circ m_1 = m_2 \circ \left(\phi \otimes \phi \right)
\hspace{1mm} \mbox{ and } \hspace{1mm}
\phi \circ \mathbb{I}_1 = \mathbb{I}_2
\end{align}
in the case of unital algebras. \\
For coalgebras ($C_1, \Delta_1$) and ($C_2, \Delta_2$), the linear map $\tilde{\phi}: C_1 \rightarrow C_2$ is an coalgebra morphism if
\begin{align}
\Delta_2 \circ \tilde{\phi} = \left(\tilde{\phi} \otimes \tilde{\phi}\right) \circ \Delta_1
\hspace{1mm}\mbox{ and }\hspace{1mm}
\hat{\mathbb{I}}_2 \circ \tilde{\phi} = \hat{\mathbb{I}}_1
\end{align}
is fulfilled. The latter only holds for the counital case.
\end{defi}
Before we come to the notion of Hopf algebras, we first need to merge algebras and coalgebras to bialgebras as described in the following definition.
\begin{defi}[Bialgebra] \label{def:bialgebra}
A $\mathbb{K}$-vector space $B$ together with a unital $\mathbb{K}$-algebra structure $(m, \mathbb{I})$ and a counital $\mathbb{K}$-coalgebra structure $\left(\Delta, \hat{\mathbb{I}}\right)$ is called a (unital and counital) $\mathbb{K}$-bialgebra $\left(B, m, \mathbb{I}, \Delta, \hat{\mathbb{I}}\right)$ if one of the following conditions hold:
\begin{itemize}
\item[(i)] The linear maps $(m, \mathbb{I})$ are morphisms of coalgebras, or
\item[(ii)] the linear maps $\left(\Delta, \hat{\mathbb{I}}\right)$ are morphisms of algebras.
\end{itemize}
\end{defi}
Note, that the requirements (i) and (ii) in the definition above are equivalent, as it was proven in \cite{Kass:QuantumGroups}. Therefore, it suffices if only one of the conditions is fulfilled.\\
Since we will always assume (co-)algebras to be (co-)unital and bialgebras to be both of it, we can conveniently waive this prefix and just refer to them as (bi-, co-)algebras. Motivated by the coproduct, there is another coassociative map one can define on bialgebras by
\begin{align}
\tilde{\Delta}&: B \rightarrow B \otimes B
\hspace{1mm}\mbox{ and } \nonumber\\
\tilde{\Delta}& \coloneqq \Delta - (\operatorname{id} \otimes \mathbb{I} + \mathbb{I} \otimes \operatorname{id}).
\label{eq:reduced coproduct}
\end{align}
The map $\tilde{\Delta}$ is called the reduced coproduct, and the space $\operatorname{Prim}(B)$ of primitive elements is given by the kernel of $\tilde{\Delta}$
\begin{align}
\operatorname{Prim}(B) \coloneqq &\ker \tilde{\Delta} \nonumber\\
&\hspace{-7mm}= \left\{b\in B: \Delta(b) = b\otimes \mathbb{I} + \mathbb{I} \otimes b\right\}.
\end{align}
Analogous to the iterated coproduct in equations (\ref{eq:iterated coproduct1}) and (\ref{eq:iterated coproduct2}), we define the iterated reduced coproduct recursively by the following definition.
\begin{gather}
\tilde{\Delta}^0 \coloneqq \operatorname{id} \hspace{1mm} \mbox{ and } \hspace{1mm} \forall m, n \in \mathbb{N}_0 , \, m\leq n : \nonumber\\
 \tilde{\Delta}^{n+1} \coloneqq \left( \operatorname{id}^{\otimes m} \otimes \tilde{\Delta} \otimes \operatorname{id}^{\otimes (n-m)}\right) \circ \tilde{\Delta}^n 
\label{eq:iterated rdeuced coproduct}
\end{gather}
since $\tilde{\Delta}$ itself is coassociative, too.
Now we will extend the notion of a bialgebra to that of a Hopf algebra.
\begin{defi}[Hopf algebra]\label{def:hopf algebra}
A Hopf algebra $\left(H, m, \mathbb{I}, \Delta, \hat{\mathbb{I}}, S\right)$ is a $\mathbb{K}$-bialgebra together with an endomorphism $S: H \rightarrow H$, called the antipode, satisfying
\begin{align} \label{eq:hopf algebra condition} \hspace{-2.3mm}
m \circ \left(S \otimes \operatorname{id}\right)\circ \Delta = \mathbb{I}\circ\hat{\mathbb{I}} = m \circ \left(\operatorname{id} \otimes S\right) \circ \Delta .
\end{align}
\end{defi}
\begin{rem}
Consider an algebra $\left(A, m, \mathbb{I}\right)$ and a coalgebra $\left(C, \Delta, \hat{\mathbb{I}}\right)$. Then, one can define an algebra $\left(\operatorname{Hom}_\mathbb{K}(C,A), \ast, e\right)$, consisting of the vector space $\operatorname{Hom}_\mathbb{K} (C,A)$ of linear maps from $C$ to $A$, a unit $e$, and a bilinear map $\ast$, called the convolution product, given by
\begin{gather}
e = \mathbb{I} \circ \hat{\mathbb{I}} 
\hspace{1mm}\mbox{ and }\hspace{1mm} \forall f, g \in \operatorname{Hom}_\mathbb{K}(C,A) :\nonumber\\
f \ast g = m \circ (f\otimes g) \circ \Delta .
\end{gather}
Taking a Hopf algebra $\left(H, m, \mathbb{I}, \Delta, \hat{\mathbb{I}}, S\right)$, the antipode $S\in \operatorname{Hom}_\mathbb{K} (H,H)$ on $H$ can be defined by
\begin{align}
S \ast \operatorname{id}_H = \operatorname{id}_H \ast S = e.
\end{align}
\end{rem}
Let $H$ be a bialgebra with antipode $S$. Then, the requirement for $H$ being a Hopf algebra can be expressed by the commutativity of the following diagram
\begin{eqnarray}
\vcenter{\hbox{
\begin{tikzpicture}
\matrix(M) [matrix of math nodes, row sep=3em, column sep=1.5em, text height=1.5ex, text depth=0.25ex]
{ & H \otimes H & & H \otimes H & \\
 H & & \mathbb{K} & & H\\
  & H \otimes H & & H \otimes H & \\} ;
\path[->]
(M-2-1) edge node[auto]{\small $\Delta$} (M-1-2)
		edge node[auto]{\small $\hat{\mathbb{I}}$} (M-2-3)
		edge node[auto,swap]{\small $\Delta$} (M-3-2)
(M-1-2) edge node[auto]{\small $S \otimes \operatorname{id}$} (M-1-4)
(M-1-4) edge node[auto]{\small $m$} (M-2-5)
(M-2-3) edge node[auto]{\small $\mathbb{I}$} (M-2-5)
(M-3-2) edge node[auto]{\small $\operatorname{id} \otimes S$} (M-3-4)
(M-3-4) edge node[auto,swap]{\small $m$} (M-2-5);
\end{tikzpicture}
}}.
\end{eqnarray}
The Hopf algebra of rooted trees, we will introduce soon, has the property to be connected and graded. Therefore, we have to clarify these terms first of all by
\begin{defi}[Connectivity and graduation]
A Hopf algebra $H$ over a field $\mathbb{K}$ is graded and connected if there exist subspaces $H_i$ such that the following conditions hold
\begin{align}
H = \bigoplus_{n \in \mathbb{N}_0} H_n, 
\hspace{2mm} H_0 \simeq \mathbb{K}, 
\hspace{2mm} H_i \equiv 0 \hspace{1mm}\forall i<0,
\end{align}
and
\begin{align}
m(H_n \otimes H_m) &= H_n H_m \subseteq H_{n+m}, \nonumber\\
\Delta H_n \subseteq \bigoplus_{i+j=n} H_i \otimes H_j &= \bigoplus_{i=0}^n H_i \otimes H_{n-i}, \\
S(H_n) &\subseteq H_n \nonumber
\end{align}
for any $n, m \in \mathbb{N}_0$.
\end{defi}

\section{Ferrers diagram}\label{app:ferrers diagram}

\begin{figure}[H]
\begin{center}
\subfigure[Ferrers diagram]{
\includegraphics[scale=0.5]{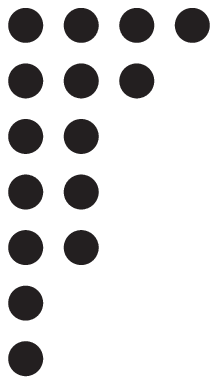}
\label{fig:ferrers diagram}
} 
\hspace{10mm}
\subfigure[Transposed Ferrers diagram]{
\includegraphics[scale=0.5]{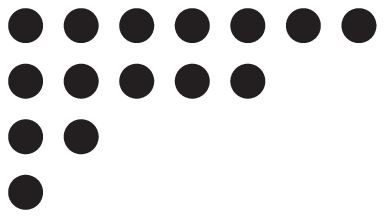}
\label{fig:ferrers diagram transposed}
}
\caption{Pictorial representation (a) of the partition $15 = 4^1 3^1 2^3 1^2$ via Ferrers diagram and (b) its conjugate partition $15 = 7^1 5^1 2^1 1^1$ as the transposed of the original diagram.}\label{fig:ferrers diagrams}
\end{center}
\end{figure}

Based upon \cite{Comtet:AdvCombinatorics} and \cite{StatonWhite:Combinatorics}, we briefly give an overview on what is called Ferrers diagram, named after the mathematician N. M. Ferrers (1829 - 1903), which can also be connected to Young diagrams.\\
Consider a partition of an integer $n$ into $k$ parts, given by the $k$-tuple $n = (y_1, \dots , y_k)$ of positive integers $y_i$ with
\begin{gather} \label{eq:app partition}
y_1 + y_2 + \dots + y_k = n \\
 \mbox{ and } \hspace{5mm} y_1 \geq y_2 \geq \dots \geq y_k \geq 1. \nonumber
\end{gather}
Alternatively, we can write $n = x_1^{m_1} \dots x_l^{m_l}$ in terms of the multiplicity $m_i$ of the different integers $x_i$ showing up in the partition, such that $x_1 > x_2 > \dots > x_l \geq 1$ and $\sum_{i=1}^l m_i = k$. For example, one possibility to decompose $n = 15$ into $k = 7$ parts is $15 = (4, 3, 2, 2, 2, 1, 1)$ or $15 = 4^1 3^1 2^3 1^2$. Note that the ladder one is not a product but a listing of the different elements in the partition and their multiplicity. \\
A useful way to represent such a partition pictorially as an array of points is Ferrers diagram. This diagram consists of $k$ rows and $y_1 = x_1$ columns, where the first row contains $y_1$ points, the second one $y_2$ points, and so on, such that the number of points in the columns and rows decreases when going from left to right and top to bottom, respectively. For example, the Ferrers diagram of the partition of $15$ into $7$ parts, we mentioned above, is shown in figure \ref{fig:ferrers diagram}. \\
There is also a conjugate partition if we consider the columns (and not the rows) of the diagram from left to right and link the number of points contained in them to positive integers $z_i$, such that $n = \sum_i z_i$. Indeed, this partition is obtained from the diagram by interchanging the rows and columns, that is to say taking the transpose of the diagram. For example, the transpose of figure \ref{fig:ferrers diagram} leads to the partition $15 = 7 + 5 + 2 + 1 = 7^1 5^1 2^1 1^1 $ (see figure \ref{fig:ferrers diagram transposed}).

\section*{Acknowledgment}
I thank Dirk Kreimer for his great supervising, strong support and encouragement.

\printbibliography

\end{multicols*}
\end{document}